\pdfoutput=1
\RequirePackage{fix-cm}
\documentclass[smallextended]{svjour3}       
\smartqed  

\usepackage{textcomp}
\usepackage{mathtools}
\usepackage{amsmath}
\usepackage{amsfonts}
\usepackage{color}
\usepackage{listings}
\usepackage{graphicx} 
\usepackage{multirow}
\usepackage[table,xcdraw]{xcolor}
\usepackage{cite}
\usepackage{paralist}
\usepackage[table,xcdraw]{xcolor}
\usepackage{siunitx}
\usepackage{rotating}
\usepackage{eqparbox}
\usepackage{todonotes}
\definecolor{whatcolor}{HTML}{d9d9d9}
\definecolor{guocolor}{HTML}{999999}

\usepackage[framemethod=tikz]{mdframed}
\usepackage{lipsum}
\usetikzlibrary{shadows}
\newmdenv[
tikzsetting= {fill=white!20},
linewidth=2pt,
roundcorner=5pt, 
shadow=true
]{myshadowbox}

\usepackage{graphics}
\usepackage{colortbl} 
\usepackage{multirow}
\usepackage{mathptmx} \usepackage[scaled=.90]{helvet} \usepackage{courier}
\usepackage{balance}
\usepackage{picture}
\usepackage[table,xcdraw]{xcolor}
\usepackage{soul}

\usepackage{array}
\usepackage{makecell}

\usepackage{verbatim}
\usepackage{algorithm}
\usepackage{algorithmicx}
\usepackage{algpseudocode}
\usepackage[export]{adjustbox}
\usepackage{mathtools}
\renewcommand{\footnotesize}{\scriptsize}
\definecolor{lightgray}{gray}{0.8}
\definecolor{darkgray}{gray}{0.6}

\usepackage[table]{xcolor}
\definecolor{Gray}{rgb}{0.88,1,1}
\definecolor{Gray}{gray}{0.85}
\definecolor{Blue}{RGB}{0,29,193}

\usepackage{booktabs}
\usepackage{xspace}
\usepackage{todonotes}

\definecolor{MyDarkBlue}{rgb}{0,0.08,0.45} 
\newcommand{\quart}[3]{\begin{picture}(100,6)
{\color{black}\put(#3,3){\circle*{4}}\put(#1,3){\line(1,0){#2}}}\end{picture}}
\newcommand{\quartex}[3]{
\begin{picture}(25,6)
    {
        \color{black}
        \put(#3,3)
        {\circle*{4}}
        \put(#1,3)
        {\line(1,0){#2}}
    }
\end{picture}
}

\definecolor{Gray}{gray}{0.95}
\definecolor{LightGray}{gray}{0.975}
\newcommand{\bi}{\begin{itemize}}
\newcommand{\ei}{\end{itemize}}
\newcommand{\be}{\begin{enumerate}}
\newcommand{\ee}{\end{enumerate}}
\newcommand{\tion}[1]{\S\ref{sect:#1}}
\newcommand{\fig}[1]{Figure~\ref{fig:#1}}

\newcommand{\eq}[1]{Equation~\ref{eq:#1}}
\newcommand{\what}{{\bf WHAT}\xspace}
\usepackage{siunitx}
\usepackage{adjustbox}

\usepackage{url}
\begin{document}

\title{Faster Discovery of Faster System Configurations with Spectral Learning}



\author{Vivek Nair \and Tim Menzies \\ \and Norbert Siegmund \and Sven Apel}


\institute{
		Vivek Nair \at
		North Carolina State University, Raleigh, USA \\
		\email{vivekaxl@gmail.com}          
           \and
		Tim Menzies \at
              	North Carolina State University, Raleigh, USA \\
		\email{tim.menzies@gmail.com}
	\and
		Norbert Siegmund\at
              	University of Weimar, Germany \\
		\email{norbert.siegmund@uni-passau.de}
	\and
		Sven Apel \at
              	University of Passau, Germany \\
		\email{apel@uni-passau.de}
}

\date{Received: date / Accepted: date}

\maketitle

\begin{abstract}
Despite the huge spread and economical importance of configurable software systems, there is  unsatisfactory support in utilizing the full potential of these systems with respect to finding performance-optimal configurations.
Prior work on predicting the performance of software configurations suffered from either (a)~requiring far too many sample configurations or (b)~large variances in their predictions.
Both these problems can be avoided using the \what spectral learner.  
{\what}'s innovation is  
the use of the spectrum (eigenvalues) of the distance matrix
between the configurations of a configurable software system, to perform dimensionality reduction. Within that
reduced configuration space, many closely associated configurations can be studied
by executing only a few sample configurations. For the subject systems studied
here, a few dozen samples yield accurate and stable predictors---less than 10\,\% prediction error, with a standard deviation of less than 2\,\%.  
When compared to the state of the art, \what (a)~requires 
2 to 10 times fewer samples to achieve similar prediction accuracies,
and (b)~its predictions are  more stable (i.e., have lower standard
deviation). 
Furthermore, we demonstrate that predictive models generated by
\what can be used by optimizers to discover system configurations that closely approach the optimal performance.
\end{abstract}

\section{Introduction}
 

Most software systems today are configurable. Despite the undeniable benefits
of configurability, 
large configuration spaces challenge developers, maintainers, and users. In the face of hundreds of configuration options, it is difficult to keep track of the effects of individual configuration options and their mutual interactions. So, predicting the performance of individual system configurations or determining the optimal configuration is often more guess work than engineering. In their recent paper, Xu et al.\ documented the  difficulties developers face
with understanding  the configuration spaces of their systems~\cite{xu2015hey}. As a result, developers tend to ignore over $5/6$ths of the configuration options, which leaves considerable optimization potential untapped and induces major economic cost~\cite{xu2015hey}.

Addressing the challenge of performance prediction and optimization in the face of large configuration spaces, researchers have developed a number of approaches that rely on sampling and machine learning~\cite{siegmund2012predicting,guo2013variability,sarkar2015cost}.
While gaining some ground, state-of-the-art approaches face two problems: 
(a)~they require far too many sample configurations for learning or (b)~they are prone to large variances in their predictions. For example, prior work on predicting performance scores using regression trees had to compile and execute hundreds to thousands of specific system configurations~\cite{guo2013variability}. 
A more balanced approach by Siegmund et al.\ is able to learn predictors for  configurable systems~\cite{siegmund2012predicting} with low mean errors, but with large variances of prediction accuracy  (e.g.\ in half of the results, the performance predictions for the Apache Web server were up to 50\,\% wrong). 
Guo et al.~\cite{guo2013variability} also proposed an incremental method to build a predictor model, which uses incremental random samples with steps equal to the number of configuration options (features) of the system. This approach also
suffered from  unstable predictions (e.g., predictions had a mean error of up to 22\,\%, with a standard deviation of up 46\,\%). Finally, Sarkar et al.~\cite{sarkar2015cost} proposed a proj\-ective-learning approach (using fewer measurements than Guo at al.\ and Siegmund et al.) to quickly compute  the number of sample configurations for learning a stable predictor. However, as we will discuss, after making that prediction, the total number of samples required for learning the predictor is comparatively high (up to hundreds of samples).

The problems of large sample sets and large variances in prediction can be avoided using the \what spectral learner, which is our main contribution.  
{\what}'s innovation is  the use of the spectrum (eigenvalues) of the distance matrix
between the configurations of a configurable system, to perform dimensionality reduction. Within that
reduced configuration space, many closely associated configurations can be studied
by measuring only a few samples.
In a number of experiments, we compared \what against the state-of-the-art approaches of Siegmund et al.~\cite{siegmund2012predicting}, Guo et al.~\cite{guo2013variability}, and Sarkar et al. \cite{sarkar2015cost} by means of six real-world configurable systems: Berkeley DB,  the Apache Web server, SQLite, the LLVM compiler, and the x264 video encoder.
We found that \what performs as well or better than prior approaches,
while  requiring far fewer samples (just a few dozen).
This is significant and most surprising, since some of the systems explored here have up to millions of possible configurations. 
Overall, we make the following contributions:
\begin{itemize}
\item We present a novel sampling and learning approach for predicting the performance of software configurations in the face of large configuration spaces. The approach is based on a
{\em spectral
learner} that uses an approximation to the first principal component of the configuration space to recursively cluster it, relying only on a few points as representatives of each cluster.
\item We demonstrate the practicality and generality of our approach by conducting experiments on six real-world configurable software systems (see Figure ~\ref{fig:systems}). The results show that our approach is more accurate (lower mean error) and more stable (lower standard deviation) than state-of-the-art approaches. A key finding is the utility of the principal component of a configuration space to  find informative samples from a large configuration space.

All materials required for reproducing this work are available at \url{https://goo.gl/689Dve}.
\end{itemize}


\begin{figure}[t]
\centering
\includegraphics[width=\columnwidth]{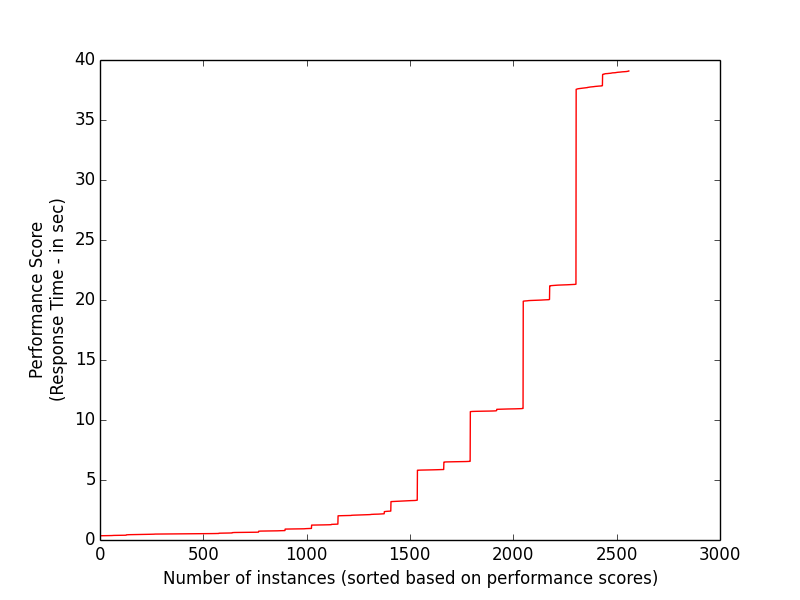}
\caption{The variation of performance scores of BDBC. }
\label{fig:motivation}
\end{figure}

\section{Background \& Related Work}  
\label{sect:addit}

We use the configurable system, BDBC, as an example to motivate our approach. BDBC is an embedded database system  written  in C. In this example, we consider 18 features or configuration options of BDBC, which the user can configure. We use the response time to indicate the performance of BDBC in different configurations. These 18 configuration options lead to 2,560 configurations. 
In Figure~\ref{fig:motivation}, we show the performance distribution of all the configurations of BDBC. It is worth noting the difference between the best performing configuration (lower left corner) and the worst performing (top right corner). The figure shows that having a good configuration reduces the response time by a factor of 40 when compared to the worst possible configuration.

An important point is that, in practice, the configurations are selected often uninformed and may not be the best or near best configuration. More over with more configurations added with ever release~\cite{xu2015hey}, it is important to have an automated approach to find the best or near-best configuration. Another aspect to this problem is the cost of evaluation (of a particular configuration), which may be very expensive. So, an ideal method should be able to find the best or near-best performing configuration with the least number of evaluations. Our approach \what{} is effective in building accurate as well as stable performance models while using fewer sample configurations than the state-of-the-art. 

A configurable software system has a set $X$ of Boolean configuration options,\footnote{In this paper, we concentrate on Boolean options, as they make up the majority of all options; see Siegmund et al., for how to incorporate numeric options~\cite{SGA+15}.} also referred to as features or independent variables in our setting.
We denote the number of features of system $S$ as $n$. The configuration space of $S$ can be represented by a Boolean space $\mathbb{Z}_{2}^{n}$, which is denoted by $F$. All valid configurations of $S$ belong to a set $V$, 
which is represented by vectors $\vec{C_i}$ (with $1\leq i\leq \left\vert{V}\right\vert$) in $\mathbb{Z}_{2}^{n}$. Each element of a configuration represents a feature, which can either be \emph{True} or \emph{False}, based on whether the feature is selected or not. 
Each valid instance of a vector (i.e., a configuration) has a corresponding performance score associated to it. 

The literature offers two approaches to performance prediction of software configurations: a {\em maximal sampling} and a {\em minimal sampling} approach: 
With {\em maximal sampling}, we compile all  possible configurations and record the associated performance scores. 
Maximal sampling  can be impractically slow. For example, the performance data used in our experiments required  26 days of CPU time for measuring (and much longer, if we also count the time required for compiling the code prior to execution). 
 Other researchers have commented that,  in 
 real world scenarios, the cost of acquiring the optimal configuration is overly expensive and time consuming \cite{weiss2008maximizing}.
 
 If collecting performance scores of all configurations is impractical,  {\em minimal sampling} 
 can be used to intelligently select and execute just enough configurations (i.e., samples) to build a
 predictive model.
 For example, Zhang et al.~\cite{zhang2015performance} approximate the
configuration space as a Fourier series, after which they can derive an expression showing how many configurations must be studied
 to build predictive models with a given error. While a theoretically satisfying result, that approach still needs thousands to hundreds of thousands of executions of sample
 configurations.  

Another set of approaches are the four "additive" {\em minimal sampling} methods of Siegmund et al.~\cite{siegmund2012predicting}.
Their first method, called feature-wise sampling ({\em FW}), is their basic method.
To explain {\em FW}, we note that, from a configurable software system, it is theoretically possible to enumerate many or all of the valid configurations\footnote{Though, in practice, this can be very difficult. For example, in models like the Linux Kernel such an enumeration is practically impossible ~\cite{sayyad13b}.}. 
Since each configuration ($\vec{C_i}$) is a vector of $n$ Booleans, it  is possible to use this information to isolate examples of how much each feature individually contributes to the total run time:
\begin{enumerate}
\item Find a pair of  configurations $\vec{C_1}$ and $\vec{C}_2$, where $\vec{C}_2$ uses exactly the same features as $\vec{C_1}$, plus one  extra feature $f_i$.
\item Set the run time $\Pi(f_i)$ for feature $f_i$ to be the difference in the performance scores between $\vec{C_2}$ and $\vec{C_1}$.
\item The run time  for a new configuration  $\vec{C}_i$ (with $1\leq i\leq \left\vert{V}\right\vert$) that has not been sampled before is then the sum of the run time of its features, as determined before:
\begin{equation}
  \Pi(C_i) = \sum_{f_j \in C_i}\Pi(f_j)  
\end{equation}
\end{enumerate}

When many pairs, such as ${\vec{C_1},\vec{C}_2}$, satisfy the criteria of point~1, Siegmund et al.\ used the 
pair that covers the {\em smallest} number of features. Their minimal sampling method, {\em FW},
compiles and executes only these smallest ${\vec{C_1}}$ and ${\vec{C_2}}$ configurations. 
Siegmund et al.\ also offers three extensions to the basic method, which are based on sampling
not just the smallest  pairs, but also additional configurations covering certain kinds of {\em interactions} between features. 
All the following minimal sampling policies compile and   execute valid configurations selected via one of three heuristics:

\begin{description}
\item[{\em PW (pair-wise):}] For each pair of features, try to find a configuration that contains the pair and has a minimal number of features selected. 
\item[{\em HO (higher-order):}] Select extra configurations, in which three features, $f_1,f_2,f_3$, are selected if two of the following pair-wise interactions exist: $(f_1,f_2)$ and $(f_2,f_3)$ and $(f_1,f_3)$.
\item[{\em HS (hot-spot):}] Select extra configurations that contain features that are
frequently interacting with other features. 
\end{description}

Guo et al.~\cite{guo2013variability} proposed a progressive random sampling approach, which samples the configuration space in steps of the number of features of the software system in question. They used the sampled configurations to train a regression tree, which is then used to predict the performance scores of other system configurations. The termination criterion of this approach is based on a heuristic, similar to the {\em PW} heuristics of Siegmund et al. 

Sarkar et al.~\cite{sarkar2015cost} proposed a cost model for predicting the effort (or cost) required to generate an accurate predictive model. The user can use this model to decide whether to go ahead and build the predictive model. This method randomly samples configurations and uses a heuristic based on feature frequencies as termination criterion. The samples are then used to train a regression tree; the accuracy of the model is measured by using a test set (where the size of the training set is equal to size of the test set). One of four projective functions (e.g., exponential) is selected based on how correlated they are to  accuracy measures. The projective function is used to approximate the accuracy-measure curve, and the elbow point of the curve is then used as the optimal sample size. Once the optimal size is known, Sarkar et al.\ uses the approach of Guo et al.\ to build the actual prediction model.

The advantage of these previous approaches is that, unlike  the results of Zhang et al., they require only dozens to hundreds of samples. Also, like our approach, they do not require to enumerate all configurations, which is important for highly configurable software systems. 
That said, as shown by our experiments (see Section~\ref{sec:experiments}), these approaches produce estimates with  larger mean errors and partially larger variances than our approach. While sometimes the approach by Sarkar et al. results in  models with (slightly)
lower mean error rates, it still requires a considerably larger number of samples (up to hundreds), while \what requires only few dozen.
 
 \section{Approach}

\subsection{Spectral Learning}\label{sect:spect}

The minimal sampling method we propose here is based on a spectral-learning algorithm
that  explores the spectrum (eigenvalues) of the distance matrix between  configurations in the configuration space.
In theory, such spectral learners are an appropriate method to handle noisy, redundant, and tightly inter-connected variables, for the following reasons:
When data sets have many irrelevancies or closely associated data parameters $d$, then
only a few eigenvectors $e$, $e \ll d$  are required to characterize the data.
In this reduced space:
\begin{itemize}
\item
Multiple inter-connected variables $i,j,k \subseteq d$ can be represented
by a single eigenvector;
\item
Noisy variables from $d$ are
ignored, because they  do not contribute to the signal in the data;
\item
Variables  become (approximately) parallel lines
in $e$ space. For  redundancies \mbox{$i,j \in d$}, we
can ignore $j$
since effects that change over $j$ also
change in the same way over $i$;
\end{itemize}
That is, in theory, samples of configurations drawn via an eigenspace sampling method
would not get confused by noisy, redundant, or tightly inter-connected variables. Accordingly,
we expect predictions built from that sample to have  lower mean errors and lower variances on that error.

Spectral methods have been used before for a variety of data mining applications~\cite{kamvar2003spectral}.
Algorithms, such as PDDP~\cite{boley98}, use spectral methods, such as principle component analysis (PCA), to
recursively divide data into smaller regions.  Software-analytics researchers use spectral methods (again, PCA) as a pre-processor prior to data mining  to reduce noise in software-related data sets~\cite{theisen2015approximating}.
However, to the best of our knowledge, spectral methods have not been used before as a basis of a minimal sampling method.

\what is somewhat different from other spectral
learners explored in, for instance, image processing applications~\cite{shi2000normalized}.
Work on image processing does not aim at
defining a minimal sampling policy to predict performance scores.
Also, a standard spectral method requires an $O(N^2)$ matrix multiplication to compute the components
of PCA~\cite{ilin10}. Worse, in the case of hierarchical division methods, such as PDDP,
the polynomial-time inference must be repeated at every level of the hierarchy.
Competitive results can be achieved
using an $O(2N)$ analysis that we have developed previously~\cite{me12d}, which is  based on  a heuristic proposed by Faloutsos and Lin~\cite{Faloutsos1995} (which Platt has shown computes a Nystr\"om approximation to the first component of PCA~\cite{platt05}).

\begin{figure}
\begin{tabular}{ccc}
  \includegraphics[height=35mm, width=35mm]{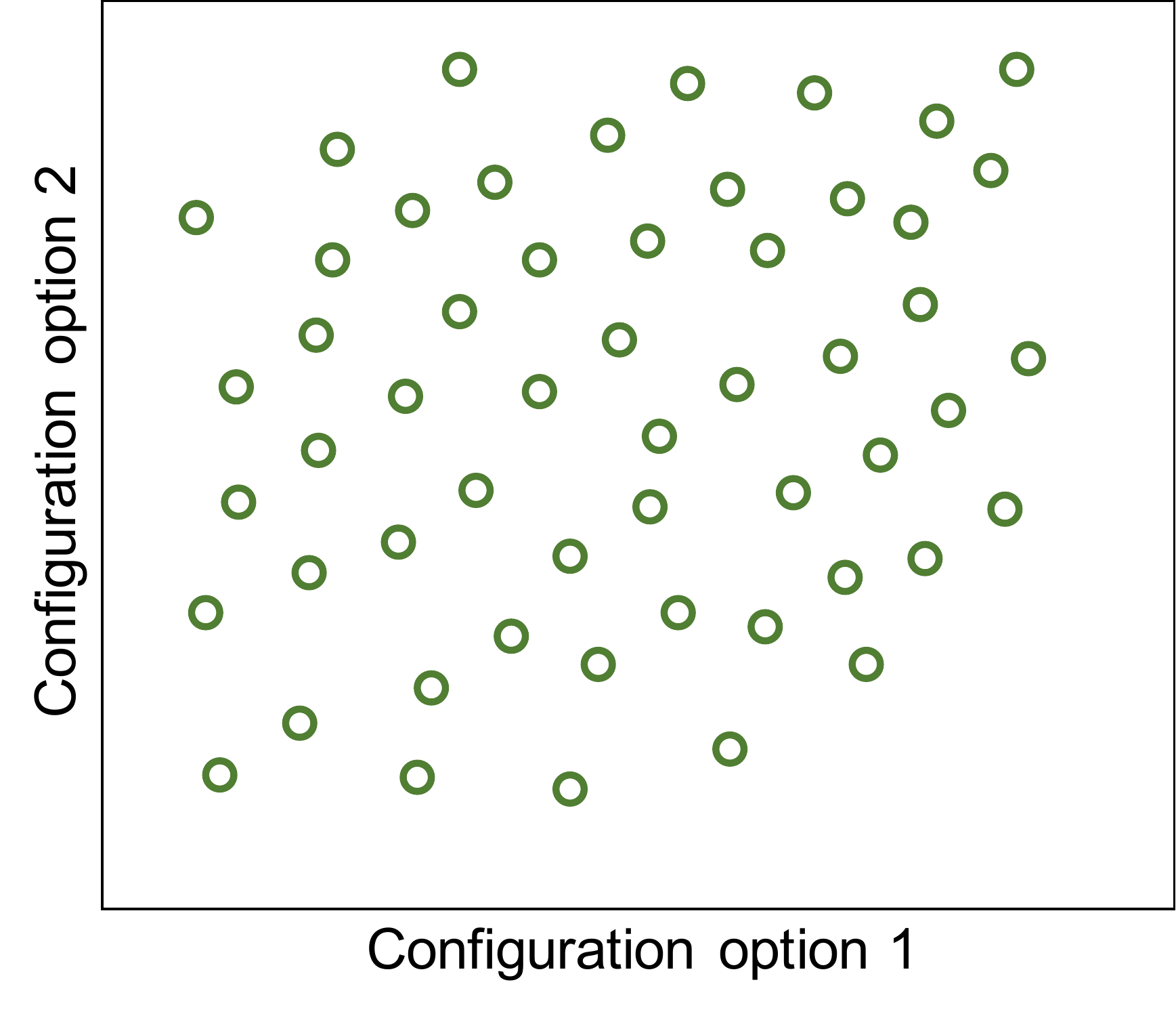} &   \includegraphics[height=35mm, width=35mm]{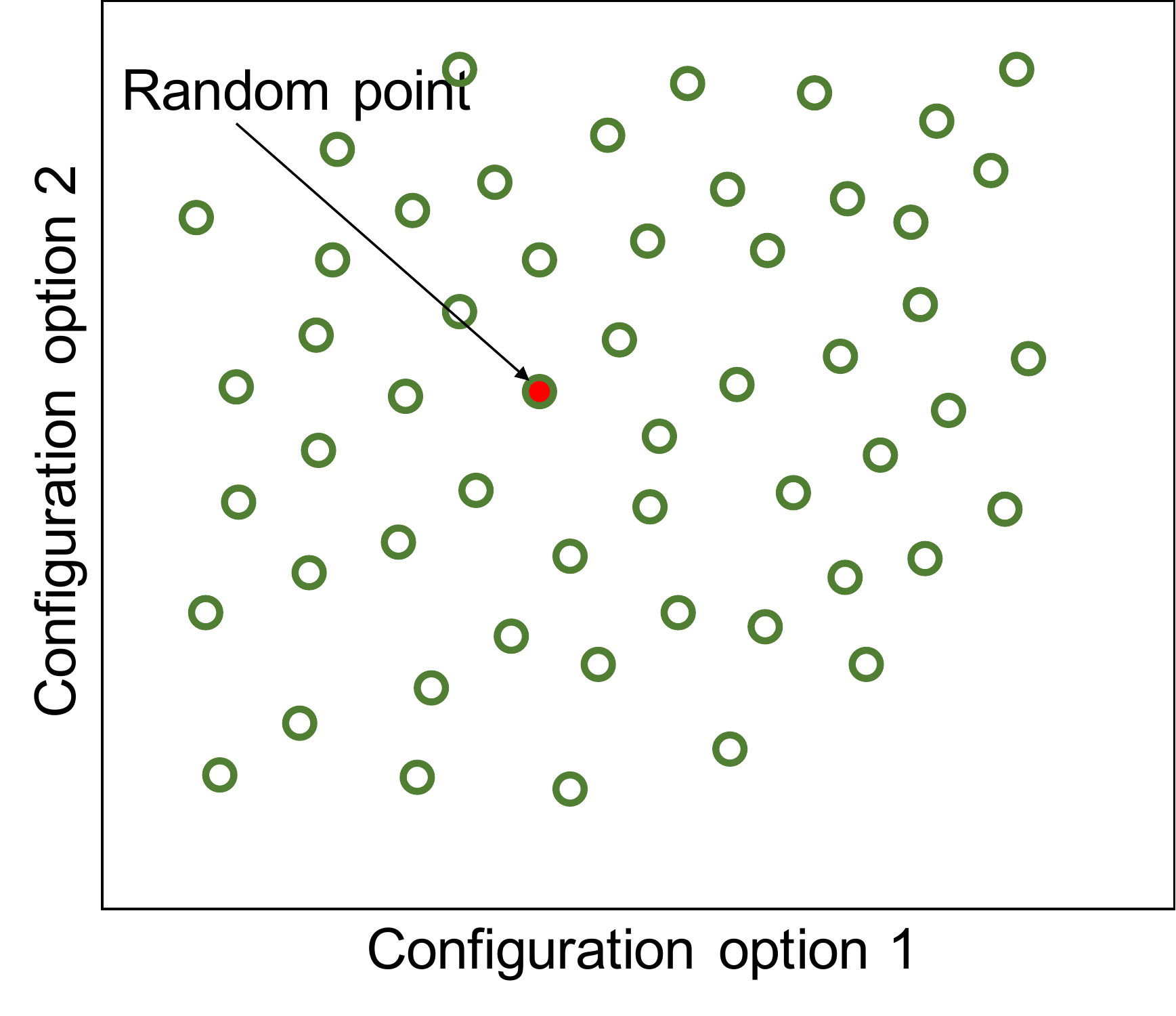}  & \includegraphics[height=35mm, width=35mm]{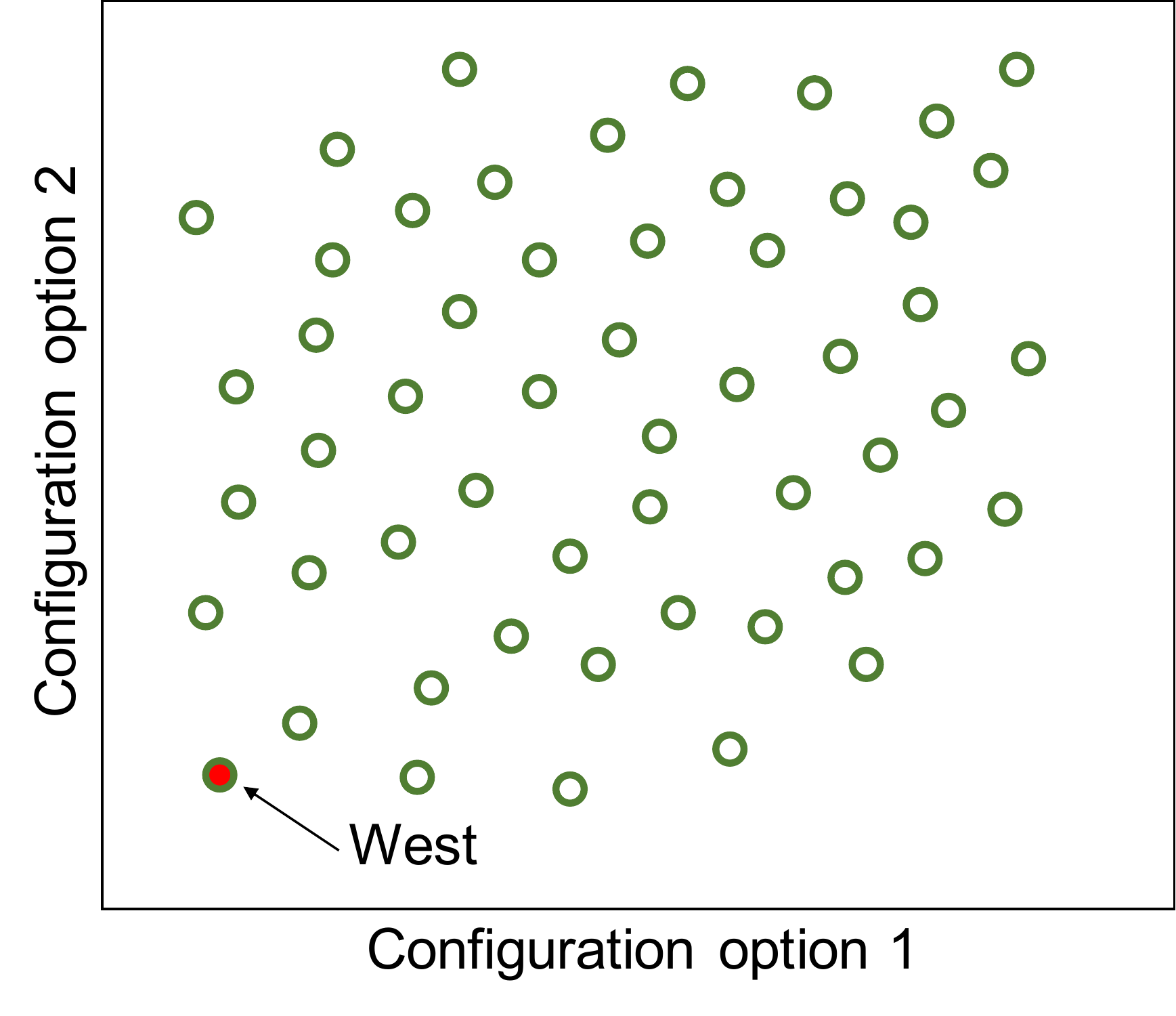}  \\ \begin{tabular}[c]{@{}l@{}}(a) Feature space of a system \\with two configurations\end{tabular}  & \begin{tabular}[c]{@{}l@{}}(b) Choosing a random point \\from the feature space\end{tabular} & \begin{tabular}[c]{@{}l@{}}(c) Find {\em West} farthest from the\\selected random point\end{tabular}\\[6pt]
 \includegraphics[height=35mm, width=35mm]{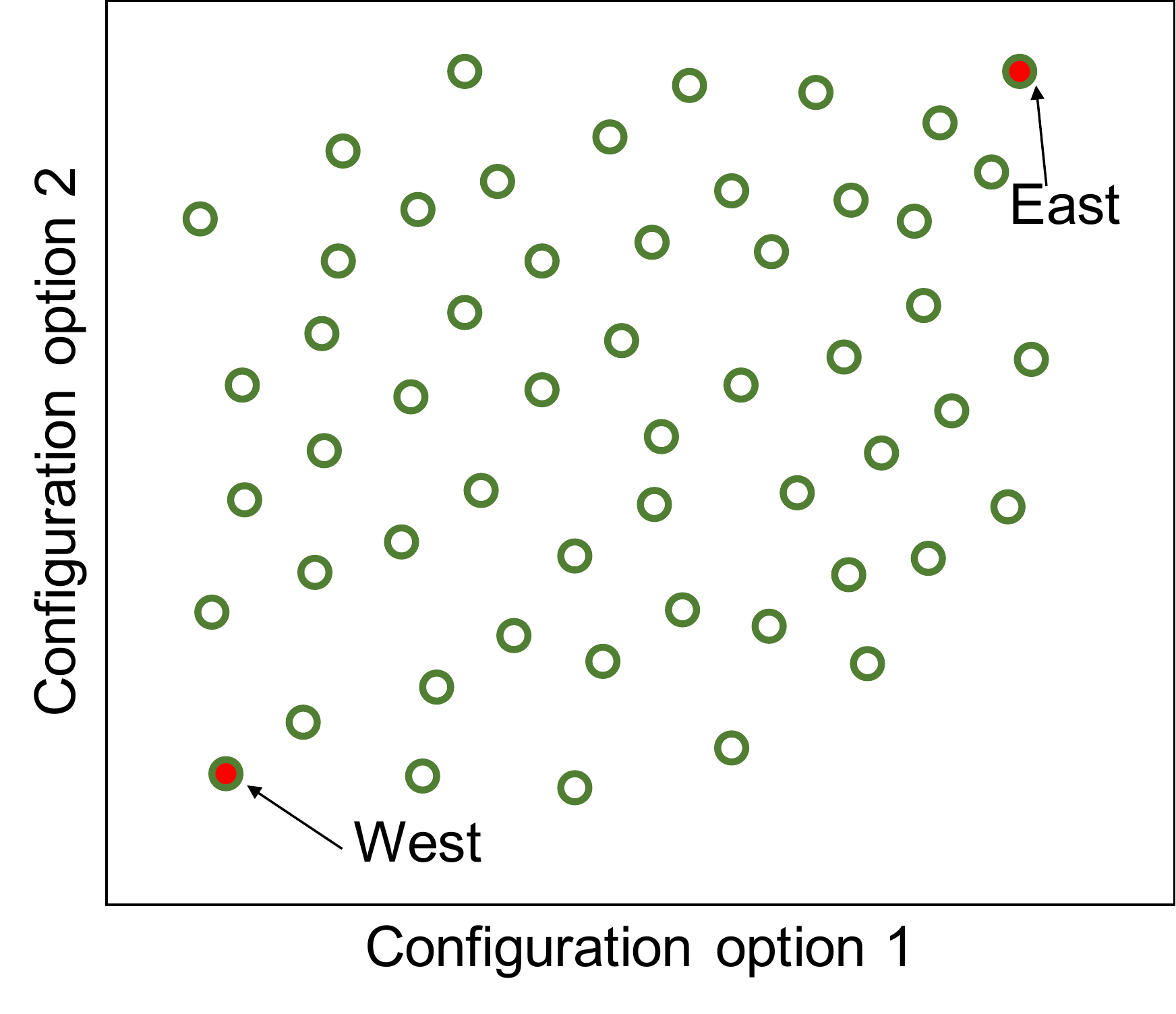} &
\includegraphics[height=35mm, width=35mm]{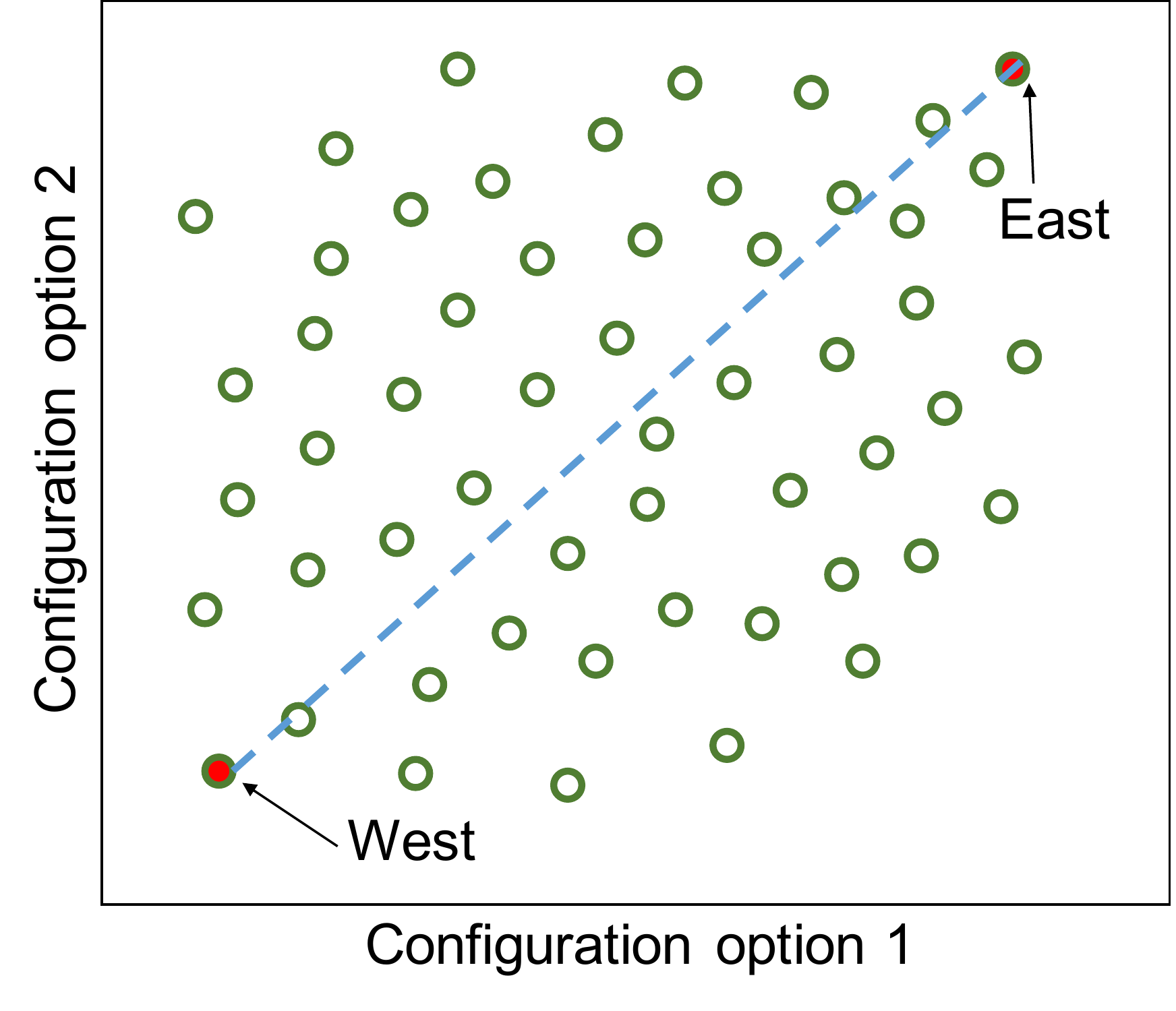} &
\includegraphics[height=35mm, width=35mm]{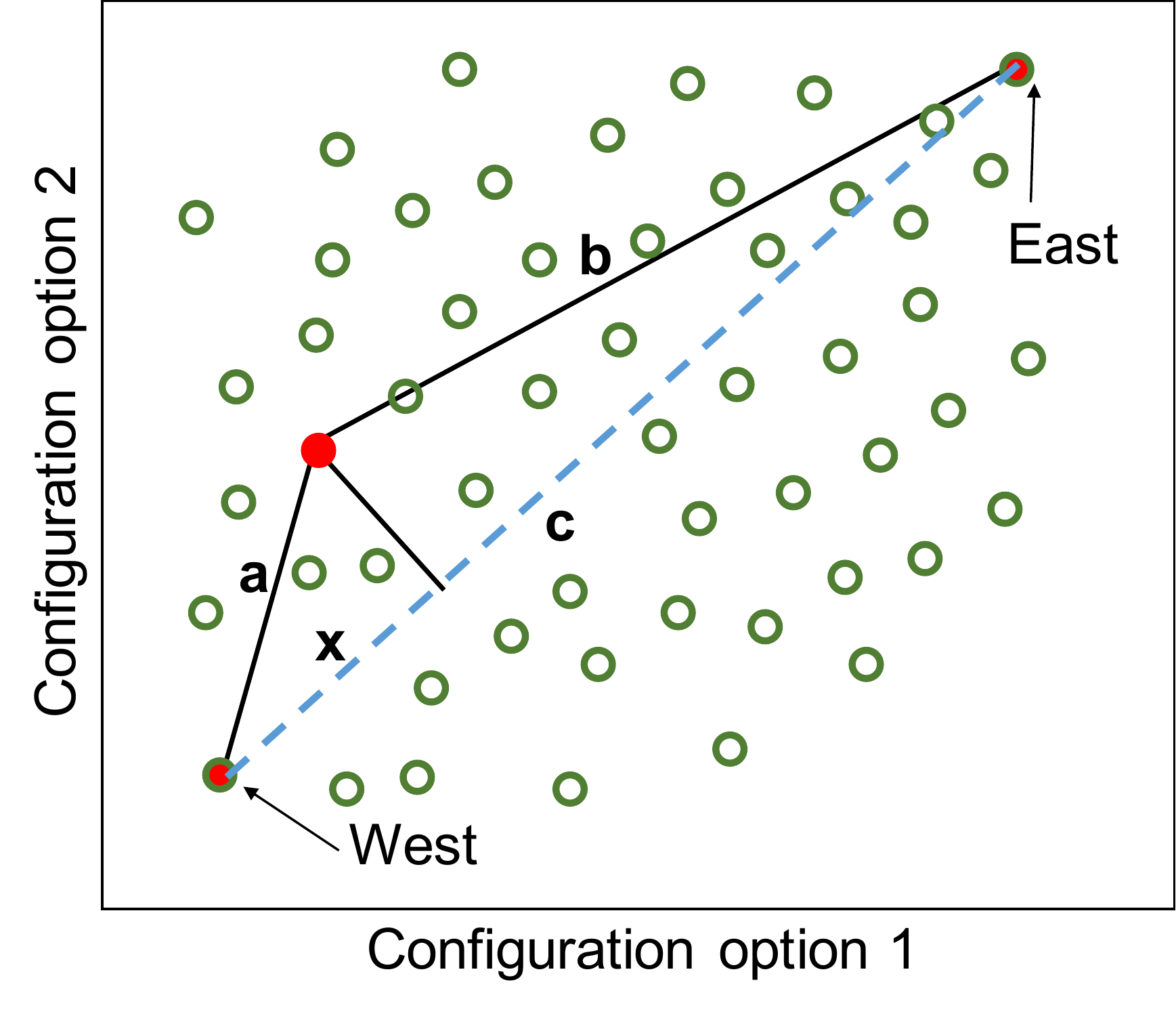} 
\\ 
     \begin{tabular}[c]{@{}l@{}}(d) Find {\em East}, a point\\farthest away from {\em West}.\end{tabular}   & \begin{tabular}[c]{@{}l@{}}(e) Line joining {\em East} and {\em West}\\is the first principal component\end{tabular} & \begin{tabular}[c]{@{}l@{}}(f) Projection of a point ({\em x})\\is calculated\end{tabular} \\[6pt]
\end{tabular}
\caption{Spectral Learning using \what{} using the  example of a system with two configuration options.}
\label{fig:spectral_desc}
\end{figure}

Figure~\ref{fig:spectral_desc} describes the procedure used to calculate the projection of a configurations. \what receives $N$ (with $1\leq \left\vert{N}\right\vert\leq \left\vert{V}\right\vert$)
valid configurations ($\vec{C}$), $N_1,N_2,...$, as input (as shown in Figure~\ref{fig:spectral_desc}(a)) and then:
\begin{enumerate}
\item
Picks any
point $N_i$ ($1\leq i \leq\left\vert{N}\right\vert$) at random (as shown in Figure~\ref{fig:spectral_desc}(b));
\item
Finds
 the point  {\em West}~$\in N$ that is
furthest away from $N_i$ (as shown in Figure~\ref{fig:spectral_desc}(c));
\item Finds the point {\em East}~$\in N$
that is furthest from {\em West} (as shown in Figure~\ref{fig:spectral_desc}(d)).
\end{enumerate}
The line joining {\em East}
and {\em West} is our approximation for the first principal component (as shown in Figure~\ref{fig:spectral_desc}(e)).
Using the distance calculation shown in Equation~\ref{eq:dist}, 
we define $\mathit{c}$ to be the distance between {\em East}~(x)
and {\em West}~(y). 
\what uses this distance ($\mathit{c}$) to divide all the configurations as follows:
The value $x_i$ is the projection of $N_i$
on the line  running  from {\em East} to {\em West} (as shown in Figure~\ref{fig:spectral_desc}(f))\footnote{The projection of $N_i$ can be calculated in the following way:\newline $a = \mathit{dist}(\mathit{East}, N_i); b = \mathit{dist}(\mathit{West}, N_i);  x_i = \sqrt{\frac{a^2 - b^2 + \mathit{c}^2}{2\mathit{c}}}$.
}.  We divide
the examples based on the median value of the projection of $x_i$. Now, we have two clusters of data divided based on the projection values (of $N_i$) on the line joining {\em East} and {\em West}. This process is applied recursively on these clusters until a predefined stopping condition. In our study, the  recursive splitting of the $N_i$'s stops when a sub-region
contains less than  $\sqrt{|N|}$ examples.
\begin{equation}
    \mathit{dist}(x, y) =     
    \begin{cases}
      \sqrt{\sum_i(x_i-y_i)^2}
      & \text{if $x_i$ and $y_i$ is numeric}\\
        \begin{cases}
            0, & \text{ if $x_i = y_i$}\\
            1, & \text{ otherwise}\\
        \end{cases}
        & \text{if $x_i$ and $y_i$ is Boolean}\\
    \end{cases}
    \label{eq:dist}
\end{equation}
We explore this approach for three reasons:
\begin{itemize}
\item
{\em It is very fast}:
This process requires only $2|n|$ distance comparisons
per level of recursion, which is far less than the $O(N^2)$
required by PCA~\cite{Du2008}
or other  algorithms such as K-Means~\cite{hamerly2010making}.
\item
{\em It is not domain-specific}:
Unlike traditional PCA, our approach is general in that it does not assume that all the variables are numeric. As shown in Equation~\ref{eq:dist},\footnote{In our study, $\mathit{dist}$ accepts pair of configuration ($\vec{C}$) and returns the distance between them. If $x_i$ and $y_i$ $\in \mathbb{R}^n$, then the distance function would be same as the standard Euclidean distance.} we can approximate distances for both numeric and non-numeric data (e.g., Boolean).

\item
{\em It reduces the dimensionality problem}:
This technique explores the underlying dimension (first principal component) without getting confused by noisy, related, and highly associated variables.
\end{itemize}

\subsection{Spectral Sampling}\label{sect:sample}
When the above clustering method terminates, our  sampling policy (which we call $S_1$) is then applied:
\begin{description}
\item[{\em Random sampling ($S_1$):}] compile and execute one  configuration,  picked at random, from each leaf cluster;
\end{description}
We use this sampling policy, because (as we will show later) it performs better than:
\begin{description}
\item[{\em East-West sampling ($S_2$):}] compile and execute the {\em East} and {\em West} poles of the leaf clusters;
\item[{\em Exemplar sampling ($S_3$):}] compile and execute all items in all leaves and return the one
with lowest performance score.
\end{description}

Note that $S_3$ is {\em not} a {\em minimal} sampling policy (since it executes all configurations). 
We use it here as one  baseline
against which we can compare the other, more minimal, sampling policies. In the results
that follow, we also compare our 
sampling methods against another baseline using information gathered after executing
all configurations.

\subsection{Regression-Tree Learning} \label{rtlearning}
After collecting the data using one of the sampling policies ($S_1$, $S_2$, or $S_3$), as described in Section \ref{sect:sample}, we  use a CART regression-tree learner~\cite{breiman1984} to build a performance predictor. Regression-tree learners seek the attribute-range split that most increases
our ability to make accurate predictions.
CART explores splits that divide $N$ samples  into two sets  $A$ and $B$, where each set  has a  standard deviation on the target variable of $\sigma_1$ and  $\sigma_2$.
CART finds the ``best'' split defined as the split that minimizes $\frac{A}{N}\sigma_1 + \frac{B}{N}\sigma_2$.
Using this best split, CART divides the data recursively.

In summary, \what  combines:
\begin{compactitem}
\item
The FASTMAP method of Faloutsos and Lin~\cite{Faloutsos1995}, which rather than $N^2$ comparisons only performs $2N$ where $N$ is the number of configurations in the configuration space;

\item A spectral-learning algorithm initially   inspired by    Boley's PDDP system~\cite{boley98}, which we modify
by replacing  PCA with FASTMAP (called
``WHERE'' in prior work ~\cite{me12d});

\item
The sampling policy that explores the leaf clusters found by this recursive division;

\item 
The CART regression-tree learner that converts the data from the samples collected by sampling policy 
into a run-time prediction model~\cite{breiman1984}.
\end{compactitem}
That is,
\begin{center}
\begin{tabular}{rcl}
WHERE& = &PDDP $-$ PCA $+$ FASTMAP\\[1.5ex] 
\what& =  & WHERE $+$ \{ $S_1, S_2, S_3$ \} $+$ CART
\end{tabular}
\end{center}
This unique combination of methods has not been previously explored in the
software-engineering literature.

\subsection{Approach as a pipeline}

Different components of \what{} can be used to sample configurations of a configurable software system, which can be then used to generate an accurate and stable performance model. We test \what{} in following way:
\begin{itemize}
    \item All the possible configurations of a system is enumerated
    \item Split the configurations into training and testing datasets based on a predefined ratio (as discussed in section~\ref{sec:exp_rig}) -- at this point none of the configurations is measured
    \item \what{} (section~\ref{sect:spect}) is used to sample configuration (section~\ref{sect:sample}) from the  training dataset and for each configuration the corresponding performance is measured.
    \item Configuration and the corresponding performance score is used to build a performance model using Regression-Tree Learner (section~\ref{rtlearning}).
    \item The accuracy (in terms of MRE) of the built performance model is measured using configurations for the testing set.
\end{itemize}
In the next section, we formulate our research questions and discuss the experimental setup.

\section{Experiments}
\label{sec:experiments}

\subsection{Research Questions} 

We formulate our research questions in terms of the challenges of
exploring large complex configuration spaces.
As our approach explores the spectral space, our hypothesis is that only a small
number of samples is required to explore the whole space.
However, a prediction model built from a very small sample of the configuration space might
be very inaccurate and unstable, that is, it may exhibit very large mean prediction errors and variances on the prediction error.

Also, if we learn models from small regions of the training data,
it is  possible that a learner will miss {\em trends} in the data
between the sample points. Such trends are useful when building {\em optimizers}
(i.e., systems that receives one configuration as input and propose an alternate
configuration that has, for instance,  a better performance). Such optimizers might
need to evaluate hundreds to millions of alternate configurations. 
To speed up that process, optimizers can use a {\em surrogate model}\,\footnote{Also known as response surface methods, meta models, or emulators.}
that  mimics the outputs of a system of interest, while being computationally cheap(er) to evaluate~\cite{loshchilov13}. For example, when optimizing
performance scores, we might ask a CART  for a performance
prediction (rather than compile and execute
the corresponding configuration).  Note that such surrogate-based
reasoning critically depends on how well the surrogate can guide optimization.

Therefore, to assess feasibility of our sampling policies, we must consider:
\begin{itemize}
\item Performance scores generated from our minimal sampling policy;
\item The variance of the error rates when comparing predicted performance scores with actual ones;
\item The optimization support offered by the performance predictor (i.e., can the model work in tandem with other off-the-shelf optimizers to generate useful solutions).
\end{itemize}

The above considerations lead to four research questions:
\begin{description}
\item[{\em RQ1:}] {\em Can  \what generate good predictions after
examining only a small number of configurations?}
\end{description}
Here, by ``good'' we mean that the predictions made by models that were trained using sampling with \what are as accurate, or more accurate,
as preditions generated from models supplied with more samples.
\begin{description}
\item[{\em RQ2:}] {\em
Do less data used in building the predictions models cause larger variances in the predicted performance scores?}
\item[{\em RQ3:}] {\em
Can ``good'' surrogate models (to be used in optimizers)
be built from minimal samples?}
\end{description}
Note that RQ2 and RQ3 are of particular concern with our approach,
since our goal is to sample as little as possible from the configuration space.
\begin{description}
\item[{\em RQ4:}] {\em How good is \what compared to the state of the art of
learning performance predictors from configurable software systems?}
\end{description}

\begin{table}
\centering
\caption{Subject systems used in the experiments.}\label{fig:systems}\footnotesize
\rotatebox{90}{
\begin{tabular}{p{1.5cm}p{2.25cm}p{0.75cm}p{0.5cm}p{6.5cm}p{1cm}p{3cm}p{1cm}}
\toprule
                     & \textbf{Description} & \textbf{LOC}     & \textbf{\#\,Feat.} & \textbf{Configurations} & \textbf{\#\,Config.}& \textbf{Benchmark Used}                                                                                                   & \textbf{Performance Metric} \\ \cmidrule{1-8}
\textbf{Apache}                                                                    & Apache is a prominent open-source Web server with numerous configuration options.                      & 230,277 & 9        & Base, HostnameLookups, KeepAlive, EnableSendfile, FollowSymLinks, AccessLog, ExtendedStatus, InMemory, Handle & 192               & We used the tools autobench and httperf to generate load on the Web server. We increased the load until the server could not, handle any, further requests & Maximum load \\ 
\\ \textbf{Berkeley~DB\newline C Edition\newline (BDBC)}    & BDBC is an embedded database system written in C.                                                                & 219,811 & 18       & HAVE\_CRYPTO, HAVE\_HASH, HAVE\_REPLICATION, HAVE\_VERIFY, HAVE\_SEQUENCE, HAVE\_STATISTICS, DIAGNOSTIC, PAGESIZE, PS1K, PS4K, PS8K, PS16K, PS32K, CACHESIZE, CS32MB, CS16MB, CS64MB, CS512MB & 2,560             & Benchmark provided by the vendor                                                                                                                           & Response time                                                \\ 
\\ \textbf{Berkeley~DB\newline Java Edition\newline (BDBJ)} &BDBJ is a complete re-development of BDBC in Java with full SQL support.                                         & 42,596  & 32       &Base, Persistence, IO, OldIO, NewIO, NIOBase, NIOType, ChunkedNIO, SingleWriteNIO, DirectNIO, LogSize, S100MiB, S1MiB, Checksum, BTreeFeatures, INCompressor, IEvictor, Evictor, Critical\_Eviction, Verifier, ITracing, Tracing, TracingLevel, Severe, Finest, Statistics  & 400               & Benchmark provided by the vendor                                                                                                                           & Response time   \\ 
\\ \textbf{LLVM}                                                                      & LLVM is a compiler infrastructure written in C++.                                                                 & 47,549  & 11       & time\_passes, gvn, instcombine, inline, jump\_threading, simplifycfg, sccp, print\_used\_types, ipsccp, iv\_users, licm & 1,024             & As benchmark, we measured the time to compile LLVM’s test suite                                                                                            & Time to\newline compile LLVM’s test suite                            \\ 
\\ \textbf{SQLite}                                                                    & SQLite is an embedded database system deployed over several millions of devices.                                 & 312,625 & 39       & OperatingSystemCharacteristics,SQLITE\_SECURE\_DELETE, ChooseSQLITE\_TEMP\_STORE,SQLITE\_TEMP\_STOREzero, SQLITE\_TEMP\_STOREone, SQLITE\_TEMP\_STOREtwo, SQLITE\_TEMP\_STOREthree, AutoVacuumOff, AutoVacuumOn, SetCacheSize, StandardCacheSize, LowerCacheSize, HigherCacheSize, LockingMode, ExclusiveLock, NormalLockingMode, PageSize, StandardPageSize, LowerPageSize, HigherPageSize, HighestPageSize & 3,932,160         & Benchmark provided by the vendor                                                                                                                           & Response time                                                \\ 
\\ \textbf{x264} & x264is a video encoder in C that provides configuration options to adjust output quality of encoded video files. & 45,743  & 16 & no\_asm, no\_8x8dct, no\_cabac, no\_deblock, no\_fast\_pskip, no\_mbtree, no\_mixed\_refs, no\_weightb, rc\_lookahead, rc\_lookahead\_20, rc\_lookahead\_40, rc\_lookahead\_60, ref, ref\_1, ref\_5, ref\_9 & 1,152 & As benchmark, we encoded the Sintel trailer (735 MB) from AVI to the xH.264 codec & Encoding time \\ \bottomrule
\end{tabular}
}
\end{table}

\noindent To answer RQ4, we will compare \what 
          against approaches presented by Siegmund et al.~\cite{siegmund2012predicting}, Guo et al.~\cite{guo2013variability}, and Sarkar et al.~\cite{sarkar2015cost}.

\subsection{Subject Systems}
\label{sec:subject_systems}
The configurable systems we used in our experiments are described in Table~\ref{fig:systems}.
All systems are real-world systems and representative of different  domains  with  different  configuration  mechanisms  and  implemented  using different programming languages.
Note, with ``predicting performance'', we 
mean predicting performance scores of the subject systems while executing test suites provided by the developers or the community, as described in Table~\ref{fig:systems}.
To compare the predictions of our and prior approaches with actual performance measures, we use data sets that have been obtained by
measuring {\em nearly all} configurations\footnote{http://openscience.us/repo/performance-predict/cpm.html}.
We say {\em nearly all} configurations, for the following reasoning: For 
all except one of our subject systems, the total number of valid configurations
was tractable (192 to 2560). However,  SQLite has 3,932,160 
possible configurations (since SQLite has 39 configuration -- $2^{39}$ possible configurations), which is an impractically large number of configurations to test whether our predictions are accurate and stable. Hence, for SQLite, we use the 4500 samples for testing prediction accuracy and stability, which we could collect in one day of CPU time. Taking this into account, we will pay particular attention to the variance of the SQLite results.

\subsection{Experimental Rig}\label{sec:exp_rig}

RQ1 and RQ2 require the construction and assessment of numerous runtime predictors from small samples
of the data. The following rig implements that construction process.

For each configurable software system, we built a table of data, one row per valid configuration. We then ran all configurations of all software systems
and recorded the performance scores (i.e., that are invoked by a benchmark).
The exception is SQLite for which we measured only the
configurations needed to detect interactions and additionally
100 random configurations.  
To this table, we added a column showing the performance score obtained from the actual measurements for each configuration.

Note that the following procedure ensures that
we \textit{never} test any prediction model on the data that we used to learn this model. Next, we repeated the following procedure 20 times (the figure of 20 repetitions was
selected using the Central Limit Theorem): 
For each subject system in \{BDBC, BDBJ, Apache, SQLite, LLVM, x264\}
\begin{itemize}
\item Randomize the order of the rows in their table of data;
\item For $X$ in \{10, 20, 30, ... , 90\};
\begin{itemize}
\item Let {\em Train} be the first $X$\,\% of the data 
\item Let {\em Test} be the rest of the data;
\item Pass {\em Train} to \what to select   sample   configurations;
\item Determine the performance scores associated with these configurations. This corresponds to a table lookup, but would entail compiling and executing a system configuration in a practical setting.
\item Using the {\em Train}  data and their performance scores, build a performance predictor using CART.
\item Using the {\em Test} data, assess the accuracy of the predictor using the error 
measure of \eq{err} (see below).
\end{itemize}
\end{itemize}

The validity of the predictors built by CART is verified on testing data. 
For each  test item, we determine how long it {\em actually} takes to run the corresponding system configuration and compare the actual measured performance to the {\em prediction} from CART. The resulting prediction error is then computed using:
\begin{equation}\label{eq:err}
\mathit{error}=\frac{\mid\mathit{predicted} - \mathit{actual}\mid}{\mathit{actual}} \cdot 100
\end{equation}
(Aside: It is reasonable to ask why this metrics and not some of the others proposed
in the literature (e.g sum absolute residuals). In short, our results are stable
across a range of different metrics. For e.g., the results of this paper have
been repeated using sum of absolute residuals and, in those other results,
we seen the same ranking of methods; see  \url{http://tiny.cc/sumAR}).

\begin{figure}[t]
\centering
\includegraphics[width=\columnwidth]{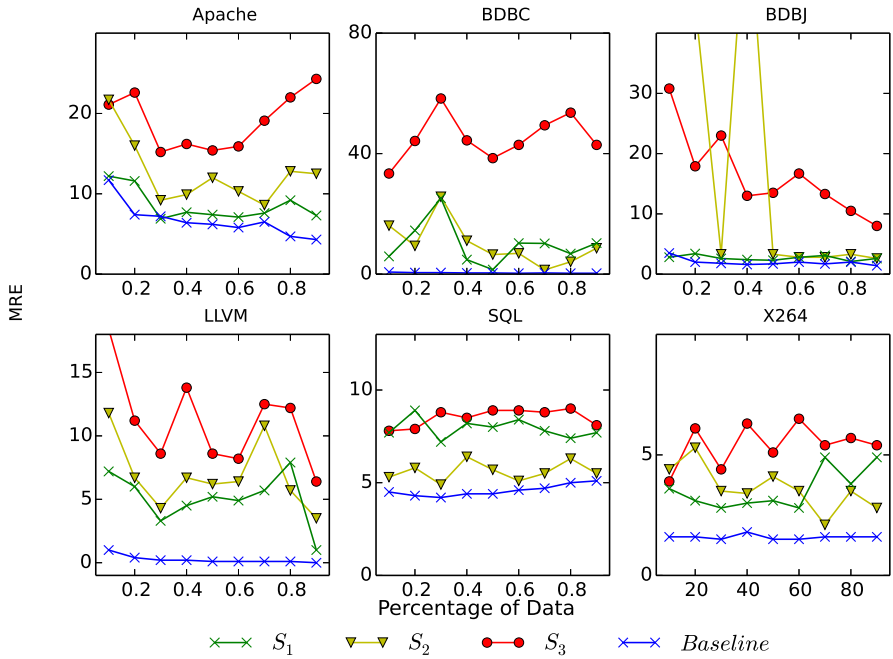}
\caption{Errors of the predictions made by \what with four different
sampling policies. Note that, on the y-axis,  {\em lower} errors are {\em better}.
}
\label{fig:sampling_accuracy}
\end{figure}

RQ2 requires testing the standard deviation of the prediction error rate. To support that test, we:
\begin{itemize}
\item Determine the $X$-th point in the above experiments, where all predictions stop improving (elbow point);
\item Measure the standard deviation of the error at this point, across our 20 repeats.
\end{itemize}
As shown in Figure~\ref{fig:sampling_accuracy}, all our results plateaued after studying $X=40$\,\% of the valid configurations\footnote{Just to clarify one frequently asked question about this work, we note
that our rig ``studies'' 40\,\% of the data. We do not mean that our predictive models
 require accessing the performance scores from the 40\,\% of the data. Rather, by ``study'' we mean   reflect 
 on a sample of configurations to determine what minimal subset of that
sample deserves to be compiled and executed.}.
 Hence to answer { RQ2}, we will compare all 20 predictions at $X=40$\,\%.
 
{ RQ3}   uses the learned regression tree as a {\em surrogate model} within an optimizer; 
\bi
\item Take   $X=40\,\%$ of the configurations;
\item Apply \what to build a CART model using some minimal sample taken from that 40\,\%;
\item Use that CART model within some standard optimizer while searching for 
configurations with least runtime;
\item  Compare the faster configurations found in this manner with the fastest configuration
known for that system.
\ei
This last item requires access to a ground truth of performance scores for a  
large number of configurations. For this experiment, we have access to that ground truth
(since we have access to all system configurations, except for SQLite). Note that such a ground truth
would not be needed when practitioners choose to use \what in their own work (it is only for our empirical investigation).

For the sake of completeness, we explored
a range of optimizers seen in the   literature:  DE~\cite{storn1997differential}, NSGA-II~\cite{deb00afast},
and our own GALE~\cite{krall2014gale,zuluaga2013active} system.   Normally,
it would be  reasonable to ask
why we used those three, and not the hundreds of other 
optimizers described in the literature~\cite{fletcher13,harman12}. However,
as shown below, all these optimizers in this
domain exhibited  very similar
behavior (all found configurations close to the
best case performance). Hence, the specific
choice of optimizer is not a critical
variable in  our analysis.

\section{Results}
\subsection{RQ1}

\begin{center}
{\em Can  \what generate good predictions after
examining only a small number of configurations?}
\end{center}

\noindent \fig{sampling_accuracy} shows the mean errors of the predictors learned
after taking $X$\,\% of the configurations, then asking  \what and some sampling method ($S_1$, $S_2$, and $S_3$)
to (a)~find what configurations to measure; then (b)~asking CART to build a predictor
using these measurements. The horizontal axis of the plots shows what $X$\,\%
of the configurations are studied; the vertical axis shows the mean relative error ($\mu$) from \eq{err}.
In this figure:
\begin{itemize}
\item
The $\times$\hspace{-2pt}---\hspace{-2pt}$\times$ lines in \fig{sampling_accuracy} show a {\em baseline} result
where data from the performance scores of 100\,\% of  configurations were used by CART
to build a runtime predictor.
\item
The other lines show the results using the sampling methods defined in Section~\ref{sect:sample}.
Note that these sampling methods used  runtime data only from a
subset of 100\,\% of the performance scores seen in configurations
from 0 to X\,\%.
\end{itemize}


In \fig{sampling_accuracy}, {\em lower} y-axis values  are {\em better} since this means lower
prediction errors. Overall, we find that:
\begin{itemize}

\item Some of the subject systems exhibit large variances in their error rate, below $X=40$\,\% (e.g., BDBC and BDBJ).
\item Above $X=40$\,\%, there is little effect on the overall change of the sampling methods.
\item
Mostly, $S_3$ shows the highest overall error, 
so that it cannot be recommended.
\item Always, the   $\times$\hspace{-2pt}---\hspace{-2pt}$\times$ baseline shows the lowest errors, which is to be
expected since predictors built on the baseline have access to all data.
\item
We see a trend that the error of  $S_1$ and $S_2$ are within $5$\,\% of the {\em baseline} results.
Hence, we can recommend these two minimal sampling methods.
\end{itemize}

\begin{figure}[t]
\centering
\includegraphics[width=0.75\columnwidth]{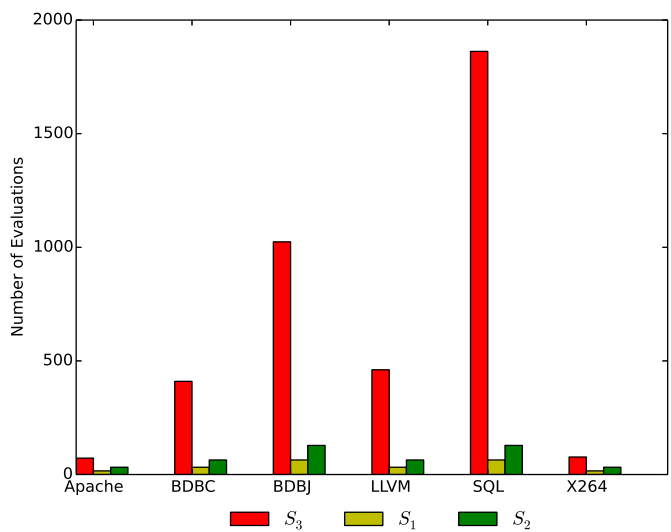}
\caption{Comparing evaluations of different sampling policies. We see that the number of configurations evaluated for $S_2$ is twice as high as $S_1$, as it selects 2 points from each cluster, where as  $S_1$ selects only 1 point. }\label{fig:Evaluations}
\end{figure}

\fig{Evaluations} provides information about which  of    $S_1$ or $S_2$ we should recommend.
This figure displays data taken from the $X=40$\,\% point of \fig{sampling_accuracy} and displays
how many performance scores of configurations are needed by our sampling methods (while
reflecting on the configurations seen in the range $0\le X \le 40$). Note that:
\begin{itemize}
\item
$S_3$ needs up to thousands of performance scores points, 
so it cannot be recommended as minimal-sampling policy;
\item $S_2$ needs twice as much performance scores as 
$S_1$ ($S_2$ uses {\em two} samples per leaf cluster  while
$S_1$ uses only {\em one}).
\item $S_1$ needs performance scores only for a few dozen (or less) configurations to generate
the predictions with the lower errors seen in \fig{sampling_accuracy}.
\end{itemize}
Combining the results of \fig{sampling_accuracy} and \fig{Evaluations}, we conclude that:

\begin{myshadowbox}
$S_1$ is our preferred spectral sampling method. Furthermore,
the answer to RQ1 is ``yes'', because applying \what{}, we can (a)~generate runtime predictors
using just a few dozens of sample performance scores; 
and (b)~these predictions have error rates
within 5\,\% of the error rates seen if predictors are built from information about all performance scores.
\end{myshadowbox}

\subsection{RQ2}

\begin{center}
{\em
Do less data used in building prediction models cause larger variances in the predicted values?}
\end{center}

\begin{figure}[t]
\includegraphics[width=\columnwidth, height=10cm]{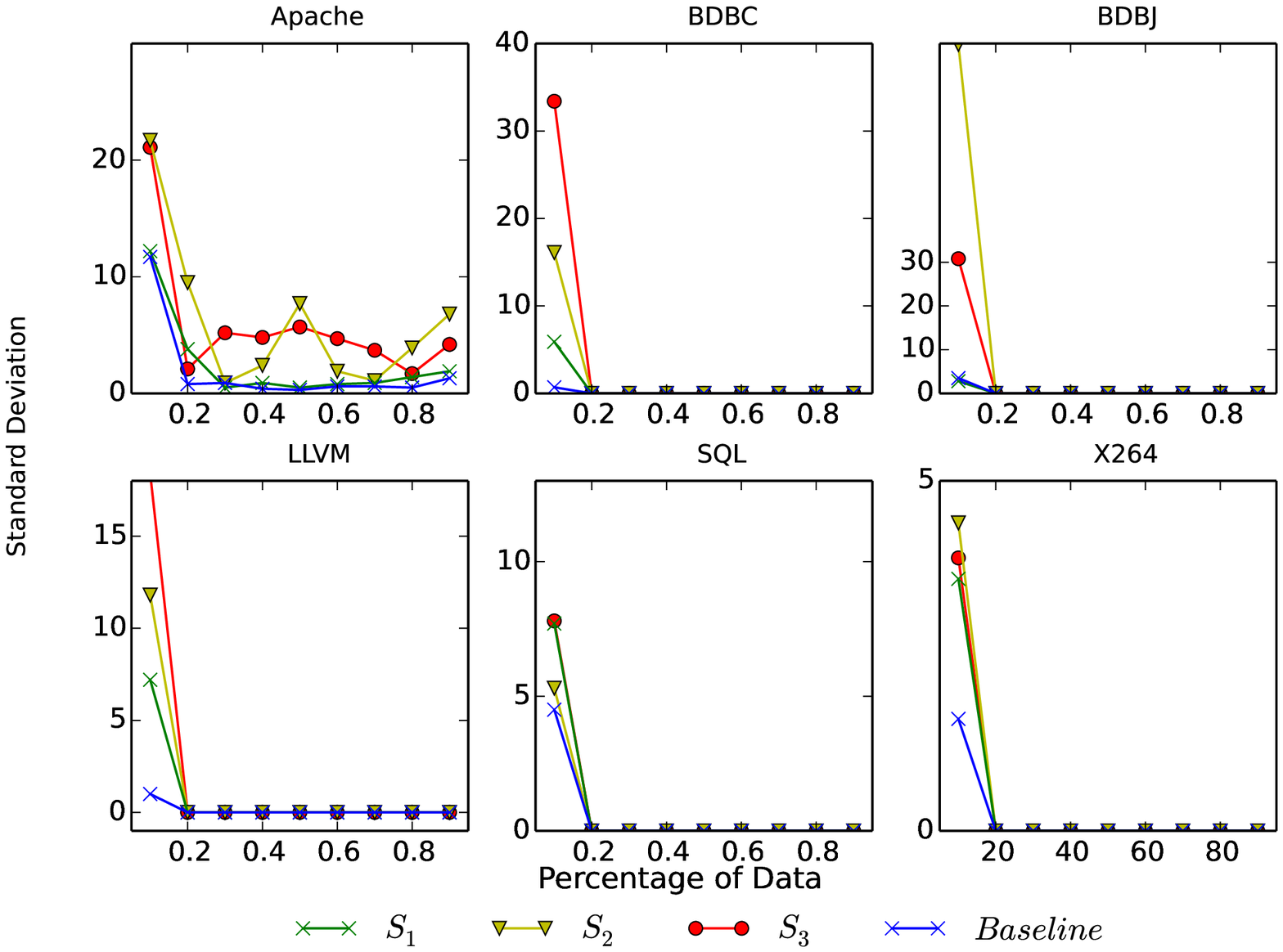}
\centering
\caption{Standard deviations seen at various points of  \fig{sampling_accuracy}.}\label{fig:Variance}
\end{figure}

Two competing effects can cause increased or decreased  variances in 
performance predictions. In our study, we report standard deviation ($\sigma$) as a measure of variances in the performance predicitons.
The   less we sample the configuration space,
the less we constrain model generation in that space. Hence, one effect that can be expected
is that models learned
from too few samples exhibit large variances. 
But,
a  compensating effect can be introduced by sampling from the spectral space
since that space contains fewer confusing or correlated variables than the raw configuration space.
\fig{Variance} reports which one of these two competing effects are dominant. 
\fig{Variance} shows that after some initial fluctuations,
after seeing $X=40$\,\% of the configurations, the variances in prediction errors reduces to nearly zero, which is similar to the results in figure \ref{fig:sampling_accuracy}.

\begin{myshadowbox}
Based of the results of Figure~\ref{fig:Variance}, we answer RQ2 with ``no'': Selecting a small number of samples does not necessarily increase variance (at least to say, not in this domain).
\end{myshadowbox}

\subsection{RQ3}

\begin{center}
{\em
Can ``good'' surrogate models (to be used in optimizers)
be built from minimal samples?}
\end{center}

The results of answering RQ1 and RQ2 suggest to use \what~(with $S_1$) to build runtime predictors from a small sample of  data. RQ3
asks if that predictor can be used by an optimizer to infer what {\em other} configurations correspond to system configurations with fast performance scores.
To answer this question,  we ran  a random set of 100 
configurations, 20 times, and related that baseline to three optimizers (GALE~\cite{krall2014gale}, DE~\cite{storn1997differential} and  NSGA-II~\cite{deb00afast}) using their
default parameters.
 
When these three optimizers mutated existing configurations to suggest new ones,
these mutations were checked for validity. Any mutants that violated the system's constraints (e.g., a feature excluding another feature) were rejected
and the survivors were ``evaluated'' by asking the CART surrogate model.
These evaluations either rejected the mutant or used it in generation $i+1$, as the basis for a search for more, possibly
better  mutants.

\begin{figure}[tb]
\centering
\includegraphics[width=\columnwidth, height=8cm]{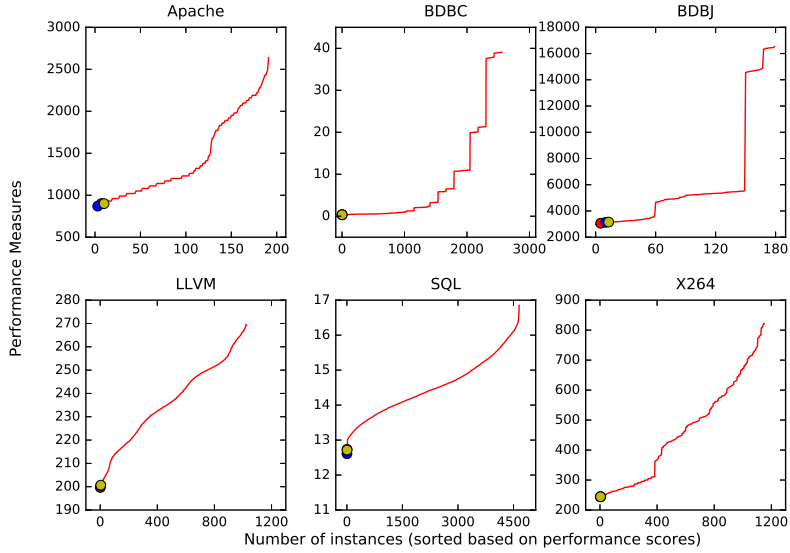}
\caption{Solutions found by GALE, NSGA-II, and DE (shown as points) laid against the ground truth (all known configuration performance scores). 
It can be observed that all the optimizers can find the configuration with  lower performance scores.}\label{fig:performance_graph}
\end{figure}

\fig{performance_graph} shows the configurations found by the three optimizers projected onto the ground truth of the performance scores of nearly
all configurations (see Section~\ref{sec:subject_systems}). Again note that, while we use that ground truth for the validation of these results, our optimizers 
used only a small part of that ground-truth data in their search for the fastest configurations (see the \what + $S_1$
results of \fig{Evaluations}).

The important information of \fig{performance_graph} is that all the optimized configurations fall within 1\,\% of the fastest
configuration according to the ground truth (see all the left-hand-side dots on each plot). Table~\ref{fig:external_validity} compares the performance of the optimizers
used in this study. Note that the performances are nearly identical, which leads to the following conclusions:

\begin{myshadowbox}
Based on the results of figure~\ref{fig:performance_graph} answer to RQ3 is ``yes'': For optimizing performance scores, we can use surrogates built from few runtime samples. The choice of the optimizer does not critically effect this conclusion.
\end{myshadowbox}

\begin{table}[tbh]
\centering
\caption{The table shows how the minimum performance scores as found by the learners GALE, NSGA-II, and DE, vary over 20 repeated
runs. Mean values are denoted $\mu$ and IQR denotes the 25th--75th percentile. A low IQR suggests that the surrogate model build by \what is stable and can be utilized by off the shelf optimizers to find performance-optimal configurations.
}
\label{fig:external_validity}
\vspace{2ex}
\begin{tabular}{lrrrrrr}
\toprule
\multirow{3}{*}{} & \multicolumn{6}{c}{Searcher}                                                                       \\ \cmidrule{2-7} 
                                  & \multicolumn{2}{c}{GALE} & \multicolumn{2}{c}{DE} & \multicolumn{2}{c}{NSGAII} \\ \cmidrule{2-7} 
                                  & Mean    & IQR    & Mean   & IQR   & Mean     & IQR     \\ \midrule
\textbf{Apache}                   & 870              & 0               & 840             & 0              & 840               & 0                \\ 
\textbf{BDBC}                     & 0.363            & 0.004           & 0.359           & 0.002          & 0.354             & 0.005            \\ 
\textbf{BDBJ}                     & 3139             & 70              & 3139            & 70             & 3139              & 70               \\ 
\textbf{LLVM}                     & 202              & 3.98            & 200             & 0              & 200               & 0                \\ 
\textbf{SQLite}                   & 13.1             & 0.241           & 13.1            & 0              & 13.1              & 0.406            \\ 
\textbf{X264}                     & 248              & 3.3             & 244             & 0.003          & 244               & 0.05             \\ \bottomrule
\end{tabular}
\end{table}


\begin{figure*}[h]
 \begin{minipage}{4in}

        {\small
          \begin{tabular}{l@{~~~~}l@{~~~~}r@{~~~~~}r@{~~~~~}c@{}r}

  \multicolumn{1}{l}{Rank}& Approach & Mean MRE($\mu$) & STDev($\sigma$) & \textbf{} & \#Evaluations \\ 
 \rowcolor{lightgray}\arrayrulecolor{lightgray}
\textbf{Apache}  & \textbf{} & \textbf{} & \textbf{}&\textbf{}&  \\\hline
  1 &       Sarkar &    7.49  &  0.82 & \quart{6}{3}{7} & 55 \\
  1 &      Guo(PW) &    10.51  &  6.85 & \quart{3}{33}{22} & 29 \\
  1 &     Siegmund &    10.34  &  11.68 & \quart{0}{55}{21} &  29\\
  1 &         	\textbf{WHAT} &    10.95  &  2.74 & \quart{16}{13}{24} & 16 \\
  1 &      Guo(2N) &    13.03  &  15.28 & \quart{7}{72}{34} &  18\\
\hline  
\rowcolor{lightgray}\arrayrulecolor{lightgray}
\textbf{BDBC}  & \textbf{} & \textbf{} & \textbf{}& \textbf{}&\\\hline
  1 &       Sarkar &    1.24  &  1.46 & \quart{0}{1}{0} &  191\\
\hline  2 &     Siegmund &    6.14  &  4.41 & \quart{4}{5}{6} &  139\\
  2 &         	\textbf{WHAT} &    6.57  &  7.40 & \quart{4}{9}{7} &  64\\
  2 &      Guo(PW) &    10.16  &  10.6 & \quart{2}{13}{11} &  139\\
\hline  3 &      Guo(2N) &    49.90  &  52.25 & \quart{16}{63}{59} &  36\\
\hline  
\rowcolor{lightgray}\arrayrulecolor{lightgray}
\textbf{BDBJ}  & \textbf{} & \textbf{} & \textbf{}& \textbf{}&\\\hline
  1 &      Guo(2N) &    2.29  &  3.26 & \quart{0}{29}{9} &  52\\
  1 &      Guo(PW) &    2.86  &  2.72 & \quart{2}{25}{14} &  48\\
  1 &         	\textbf{WHAT} &    4.75  &  4.46 & \quart{12}{40}{31} &  16\\
\hline  2 &       Sarkar &    5.67  &  6.97 & \quart{6}{62}{39} &  48\\
  2 &     Siegmund &    6.98  &  7.13 & \quart{16}{63}{51} &  57\\
\hline  
\rowcolor{lightgray}\arrayrulecolor{lightgray}
\textbf{LLVM}  & \textbf{} & \textbf{} & \textbf{}&\textbf{}& \\\hline
  1 &      Guo(PW) &    3.09  &  2.98 & \quart{0}{21}{10} &  64\\
  1 &         	\textbf{WHAT} &    3.32  &  1.05 & \quart{9}{7}{12} &  32\\
  1 &       Sarkar &    3.72  &  0.45 & \quart{13}{3}{15} &  62\\
  1 &      Guo(2N) &    4.99  &  5.05 & \quart{11}{36}{24} &  22\\
\hline  2 &     Siegmund &    8.50  &  8.28 & \quart{21}{58}{49} &  43\\
\hline

\rowcolor{lightgray}\arrayrulecolor{lightgray}
\textbf{SQLite} & \textbf{} & \textbf{} & \textbf{} & \textbf{}&\\\hline
  1 &       Sarkar &    3.44  &  0.10 & \quart{0}{0}{0} &  925\\
\hline  2 &         	\textbf{WHAT} &    5.60  &  0.57 & \quart{7}{2}{8} &  64\\
\hline  3 &      Guo(2N) &    8.57  &  7.30 & \quart{2}{28}{19} &  78\\
  3 &      Guo(PW) &    8.94  &  6.24 & \quart{6}{24}{20} &  566\\
\hline  4 &     Siegmund &    12.83  &  17.0 & \quart{16}{63}{35} &  566\\
\hline  
\rowcolor{lightgray}\arrayrulecolor{lightgray}
\textbf{x264} & \textbf{} & \textbf{} & \textbf{}& \textbf{}&\\\hline
  1 &       Sarkar &    6.64  &  1.04 & \quart{4}{2}{5} &  93\\
  1 &         	\textbf{WHAT} &    6.93  &  1.67 & \quart{4}{4}{6} &  32\\
  1 &      Guo(2N) &    7.18  &  7.07 & \quart{0}{15}{6} &  32\\
  1 &      Guo(PW) &    7.72  &  2.33 & \quart{4}{5}{8} &  81\\
\hline  2 &     Siegmund &    31.87  &  21.24 & \quart{32}{47}{61} &  81\\
\hline 

  \end{tabular}} 
\end{minipage}
\caption{Mean MRE($\mu$) seen in 20 repeats. Mean MRE is the prediction error as described in Equation~\ref{eq:err} and STDev ($\sigma$) is the standard deviation of the MREs found during multiple repeats. 
Lines with a a dot in the middle 
(e.g. \protect \quartex{3}{13}{13}) 
show the mean as a round dot withing the IQR (and if the IQR is very small, only a  round dot will be visible). 
All the results are sorted by the mean values: a lower mean value of MRE is better than large mean value. 
The left-hand side column ({\textit rank}) ranks  various techniques for e.g. when comparing various techniques for Apache, all the techniques have the same rank since their mean values are not statistically different. \textit{Rank} is computed using Scott-Knott, bootstrap 95\% confidence, and A12 test.
}
\label{fig:stats}
\end{figure*}

\subsection{RQ4}


 \begin{center}
{\em How good is \what compared to the state of the art of learning performance predictors from configurable software systems?}
\end{center}

We compare \what with the three state-of-the-art predictors proposed in the literature~\cite{siegmund2012predicting}, \cite{guo2013variability}, \cite{sarkar2015cost}, as discussed in Section~\ref{sect:addit}. Note that all approaches use regression-trees as predictors, except Siegmund's approach, which uses a regression function derived using linear programming.
The results were studied using non-parametric tests, which was also used by Arcuri and Briand at ICSE
'11~\cite{mittas13}). For testing statistical significance,
we used non-parametric bootstrap test 95\% confidence~\cite{efron93} followed by
an A12 test to check that any observed differences were not trivially small effects;
i.e. given two lists $X$ and $Y$, count how often there are larger
numbers in the former list (and there there are ties, add a half mark):
$a=\forall x\in X, y\in Y\frac{\#(x>y) + 0.5*\#(x=y)}{|X|*|Y|}$
(as per Vargha~\cite{Vargha00}, we say that a ``small'' effect has $a <0.6$). 
Lastly, to generate succinct reports, we use the Scott-Knott test to recursively
divide our optimizers. This recursion used A12 and bootstrapping  
to group together subsets that are (a)~not significantly different and are (b)~not
just a small effect different to each other. This use of Scott-Knott is endorsed
by Mittas and Angelis~\cite{mittas13}
and by Hassan et al.~\cite{7194626}.

As seen in Figure~\ref{fig:stats}, the FW heuristic of Siegmund et al. (i.e., the sampling approach using the fewest number of configurations)  has the higher errors rate and the highest standard deviation on that error rate (four out of six times). Hence, we cannot recommend this method or, if one wishes to use this method, we recommend using the other sampling heuristics (e.g., HO, HS) to make more accurate predictions (but at the cost of much more measurements). Moreover, the size of the standard deviation of this method causes further difficulties in estimating which configurations are those exhibiting a large prediction error.

As to the approach of Guo et al. (with PW), it does not standout on any of our measurements. Its error results are within 1\% of \what; its standard deviations are usually larger and it requires much more data than \what (Evaluations column of the figure~\ref{fig:stats}). 
In terms of the number of measure samples required to build a model, the right-hand column of Figure~\ref{fig:stats} shows that \what requires the fewest samples except for two cases: the approach of Guo et al. (with 2N) working on BDBC and LLVM. In both these cases, the mean error and standard deviation on the error estimate is larger than \what. Furthermore, in the case of BDBC, the error values
 are $\mu=14\,\%$, $\sigma=13\,\%$, which are much larger
than \what{}'s error scores of $\mu=6\,\%$, $\sigma=5\,\%$. 

Although the approach of Sarkar et al. produces an error rate that is sometimes less than the one of \what, it requires the highest number of measurements. Moreover, \what\textquotesingle s   accuracy is close to Sarkar\textquotesingle s approach (1\% to 2\%) difference). Hence, we cannot recommend this approach, too.

Table~\ref{tab:measurements} shows the number of evaluations used by each approaches. We see that most state-of-the-art approaches often require many more samples than
\what{}.  Using those fewest numbers of samples, \what has
within 1\% to 2\,\% of the lowest standard deviation rates 
and within 1 to 2\,\% of lowest error rates.
The exception is Sarkar's approach, which has 5\,\% lower mean error
rates (in BDBC, see the Mean MRE column of figure~\ref{fig:stats}).  However, 
as shown in right-hand side of Table~\ref{tab:measurements}, Sarkar's approach needs nearly three times
more measurements than \what. 

\noindent To summarize, there are two cases in Figure~\ref{fig:stats} where \what performs worse than, at least, one
other method:
\begin{itemize}
\item 
SQLite: The technique proposed by Sarkar et al. does better than \what (3.44 vs 5.6)
but, as shown in the final column of Figure~\ref{fig:stats},
does so at the cost of $\frac{925}{64} \approx 15$ times more evaluations that \what.
In this case, a pragmatic engineer could well prefer our solution over that of Sarkar et al. (since
number of evaluations performed by Sarkar et al.more than an order of magnitude than \what).
\item BDBC: Here again, \what is not doing the best but, compared to the number of evaluations required by all other solutions, it  is not doing particularly bad.
\end{itemize}

\noindent Given
the overall reduction of the error is   small (5\,\% difference
between Sarkar and \what in mean error), the 
cost of tripling the data-collection cost is
often not feasible in a practical context and might not justify the small additional benefit in accuracy.

\begin{myshadowbox}
Based on the results of figure~\ref{fig:stats}, we answer {\bf RQ4} with ``yes'',
since \what yields predictions that are similar to or more accurate than prior
work, while requiring fewer samples.
\end{myshadowbox}

\begin{table}[t]
\caption{Comparison of the number of the samples
required with the state of the art. The grey colored cells indicate the approach that requires the lowest number of samples.  We notice that WHAT and Guo (2N) uses less data compared to the other approaches. The high fault rate  of Guo (2N) accompanied with high variability in the predictions makes WHAT our preferred method.}\label{tab:measurements}
\vspace{2ex}
\centering
\small
\begin{tabular}{lrrrrr}
\toprule
                                   & \multicolumn{5}{c}{{Samples}}                                                                         \\ \cmidrule{2-6} 
\multirow{-2}{*}{\textbf{}} & {Siegmund} & {Guo (2N)}          & {Guo (PW)} & {Sarkar} & {WHAT}                \\ \midrule
\textbf{Apache}                    & 29                & 181                        & 29                & 55              & \cellcolor[HTML]{C0C0C0}16 \\ 
\textbf{BDBC}                      & 139               & \cellcolor[HTML]{C0C0C0}36 & 139               & 191             & 64                         \\ 
\textbf{BDBJ}                      & 48                & 52                         & 48                & 57              & \cellcolor[HTML]{C0C0C0}16 \\ 
\textbf{LLVM}                      & 62                & \cellcolor[HTML]{C0C0C0}22 & 64                & 43              & 32                         \\ 
\textbf{SQLite}                    & 566               & 78                         & 566               & 925             & \cellcolor[HTML]{C0C0C0}64 \\ 
\textbf{X264}                      & 81                & \cellcolor[HTML]{C0C0C0}32 & 81                & 93              & \cellcolor[HTML]{C0C0C0}32 \\ \bottomrule
\end{tabular}
\end{table}

\section{Why does it work?}
In this section, we present an in-depth analysis to understand why our sampling technique (based on a spectral learner) achieves such low mean fault rates while being stable (low variance). We hypothesize that the configuration space of the system configuration lie on a low dimensional manifold. 

\subsection{History}
Menzies et. al~\cite{me12d}  demonstrated how to exploit the underlying dimension to cluster data to find local homogeneous data regions in an otherwise heterogeneous data space. The authors used an algorithm called WHERE (see section~\ref{rtlearning}), which recurses on two dimensions synthesized in linear time using a technique called FASTMAP ~\cite{Faloutsos1995}. The use of underlying dimension has been endorsed by various other researchers~\cite{bettenburg2012think, deiters2013using, bettenburg2015towards, zhang2016cross}. There are numerous other methods in the literature, which are used to learn the underlying dimensionality of the data set such as Principal Component Analysis (PCA)~\cite{jolliffe2002principal}~\footnote{WHERE is an approximation of the first principal component}, Spectral Learning~\cite{shi2000normalized} and Random Projection~\cite{bingham2001random}.  These algorithms  use  different techniques to identify the underlying, independent/orthogonal dimensions to cluster the data points and differ with respect to the computational complexity and accuracy. We use WHERE since it computationally efficient~$O(2N)$, while still being accurate.

\subsection{Testing Technique}
Given our hypothesis the configuration space lies in a lower dimensional hyperplane ---  it is imperative to demonstrate that the intrinsic dimensionality of the configuration space is less than the actual dimension. To formalize this notion, we borrow the concept of correlation dimension from the domain of physics~\cite{grassberger2004measuring}. The correlation dimension of a dataset with $k$ items is found by computing the number of items found at distance withing radius $r$ (where r is the Euclidean distance between two configurations) while varying $r$. This is then normalized by the number of connections between $k$ items to find the expected number of neighbors at distance $r$. This can be written as:
\begin{equation}
    C(r) = \frac{2}{k(k-1)} \displaystyle\sum_{i=1}^{n} \displaystyle\sum_{j=i+1}^{n} I(||x_i, x_j|| < r) \\  
 \end{equation} 
$$
where:
    I(x < y) = \begin{cases}
            1, & \text{ if x \textless y}\\
            0, & \text{ otherwise}\\
    \end{cases}
$$

Given the dataset with $k$ items and range of distances [$r_0$--$r_{max}$], we estimate the intrinsic dimensionality as the mean slope between $\ln(C(r))$ and $\ln(r)$.

\subsection{Evaluation}
\begin{figure}[t]
\includegraphics[width=0.9\columnwidth]{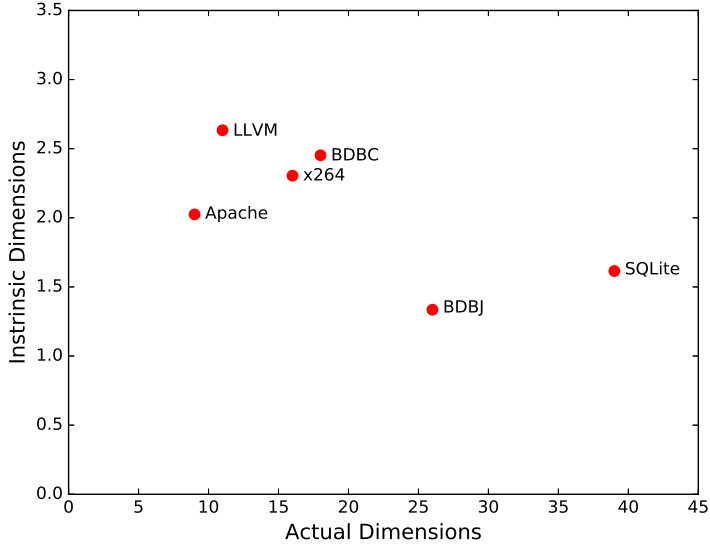}
\caption{The actual dimensions are shown on the x-axis and intrinsic dimensionality is shown on the y-axis. The points are annotated with the names of the corresponding software system. The intrinsic dimensionality of the systems are much lower than the actual dimensionality (number of columns in the dataset).}
\label{fig:underlying_d}
\end{figure}
On the configuration space of our subject systems, we observe that {the intrinsic dimensionality of the software system is much lower than the actual dimension}. Figure~\ref{fig:underlying_d} presents the intrinsic dimensionality along with the actual dimensions of the software systems. If we take a look at the intrinsic dimensionality and compare it with the actual dimensionality, then it becomes apparent that the configuration space lies on a lower dimensional hyperplane. For example, SQLite has 39 configuration options, but the intrinsic dimensionality of the space is just 1.61 (this is a fractal dimension). At the heart of \what is WHERE (a spectral clusterer), which uses the approximation of the first principal component to divide the configuration space and hence can take advantage of the low intrinsic dimensionality. 

As a summary, our observations indicate that the intrinsic dimension of the configuration space is much lower that its actual dimension. Hence, clustering based on the intrinsic dimensions rather than the actual dimension would be more effective. In other words, configurations with similar performance values lie closer to the intrinsic hyperplane, when compared to the actual dimensions, and may be the reason as to why \what achieves empirically good results.


\begin{figure}[t]
\includegraphics[width=\columnwidth]{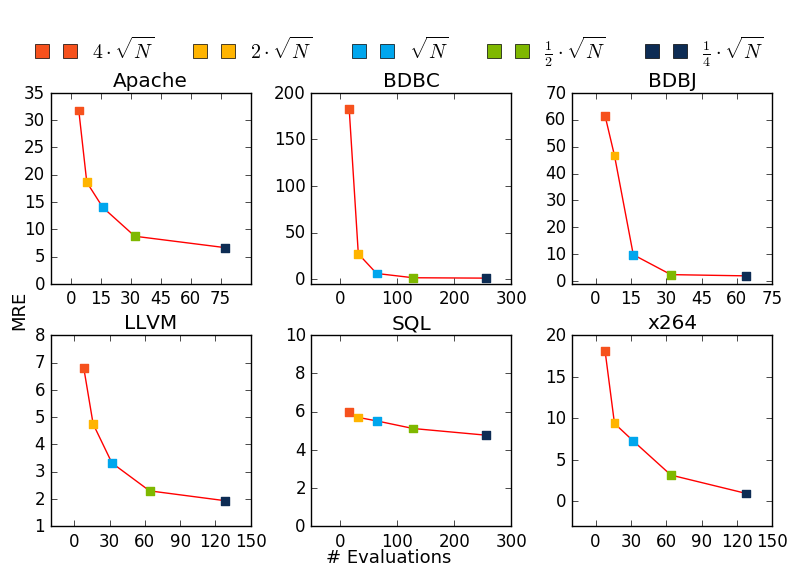}
\caption{The trade-off between the number of evaluations (affected by the size of the sub-region) and the performance (MRE) of the model generated.}
\label{fig:param_tuning}
\end{figure}

\section{Discussion}
\subsection{What is the trade-off between the MRE and the number of measurements?}
\what{} requires that the practitioner to define a stopping criterion, (size of the sub-region) before the
process commences.
The stopping criterion preempts the process of recursive division of regions
based on projection values of the configurations. In our experiments, the
number of measurements or the size of the training set depends
on the stopping criterion. An early termination of the
sampling process would lead to a very inaccurate performance model, while
late termination would result in resource wastage. Hence, it is very
important to discuss the trade-off between the upper bound of the size of the sub-region and
the MRE of the model built. In Figure~\ref{fig:param_tuning}, we show the trade-off
between the MRE found and the number
of measurements (size of training set). The trade-off characterizes
the relationship between two conflicting objectives, for example,
point in Apache, (size of sub-region=$4\cdot \sqrt{N}$) requires very few measurements but the MRE of the model built is the highest, whereas point (size of sub-region=$\frac{1}{4}\cdot \sqrt{N}$) requires large
number of measurements, but the MRE of the model built is the lowest. Since
our objective is to minimize the number of measurements while
reducing MRE, we assign the value of $\sqrt{N}$ to the upper bound of the size of the sub-region for the
purposes of our experiments.

\subsection{What is the relationship between intrinsic dimensionality and difficulty of a problem (or dataset)?}

Houle et al.~\cite{houle2012generalized} observe a clear correlation between dimensionality of a problem space and loss of performance, that is a problem represented in lower dimensions is easier to model than the same problem represented in higher dimensions. In a similar vein, Domingos~\cite{domingos2012few} explains how our intuitions fail in higher dimensions and algorithms that work on lower dimensions does not work on higher dimensions. This is because the size of the training data required to create a generalized 'model' for such a high dimensional space is exponentially large~\footnote{Another challenge of having high dimensional search space is the amount of noise induced by irrelevant dimensions.}. This is generally referred to as the ``curse of dimensionality'', but what counteracts this is the ``blessing of non-uniformity''. Blessing of non-uniformity refers to the fact that the possible valid solutions in a space is not spread uniformly across the problem space but concentrated on or near a lower dimensional manifold.  Hence, it is a rule of thumb of machine learning practitioners to reduce the dimension of a data set by projecting the data onto a lower dimensional orthogonal subspace that captures the variation in the data. Bruges~\cite{burges2010dimension} mentions that, if data lies in a lower dimensional space (with lower intrinsic dimensions), then modeling the data directly in lower dimensional manifold make it much easier to model. Our results are inline with the observation made by Bruges, as we show that few samples are enough to model a  large (sometimes millions) space, which can be attributed to the low intrinsic dimensionality of the space. 
~\\
There are several other techniques (similar to \what) that also exploit the non-uniformity of the data points, such as Random Projections~\cite{dasgupta2000experiments} and Auto-encoders~\cite{hinton2006reducing}. The central intuition  is similar to our work: problems that contain    intrinsic lower dimensions should be   easier/cheaper to model than those with higher intrinsic dimensionality. 
That said, to the best of our knowledge, we are the first to propose exploring the lower intrinsic dimensionality of  configuration spaces,  and exploit those lower dimensions  for the purposes of sampling.

\subsection{What are the limitations of WHAT?}

The limitations of \what{} are:
\begin{itemize}
    \item \what{} cannot be used for non-numeric configuration options. However, it can be used for numeric configurations options, not just Boolean options ( most related work only supports Boolean options). 
    \item The configurable systems used in this paper are fairly easy to model using machine learning techniques such as CART, but there exist software systems, which cannot be model using CART (even using 40\% of all possible configurations). For these systems, \what{} cannot be used to build accurate performance models.
    \item The effectiveness of \what{} depends on projecting the configurations on the approximated first principal component. The approximation of the first principal component require calculating the farthest points (points which is most dissimilar)  in the configuration space using euclidean distance. However, there maybe systems where  Euclidean distance (as used in this paper) cannot  find the most dissimilar points~\cite{chen2016sampling}. For such systems, \what{} in current form will not be effective (which is a part of our future work).
    \item Finding near-optimal configuration can become challenging when the configuration space is non-convex. However, we did not find such systems during our empirical evaluations. The other point we would like to stress is: we wanted to build a tool, which can differentiate between good and not-so-good configurations using few evaluations and our goal is not to find the best configuration but rather near optimal solutions.
\end{itemize}

\section{Reliability and Validity}\label{sect:construct}

{\em Reliability} refers to the consistency of the results obtained
from the research.  For example,   how well independent researchers
could reproduce the study? To increase external
reliability, we took care to either  clearly define our
algorithms or use implementations from the public domain
(SciKitLearn)~\cite{scikit-learn}. Also, all the data used in this work are available
on-line in the PROMISE\footnote{\url{http://openscience.us/repo/performance-predict/cpm.html}} code repository and all our algorithms
are on-line at github.com/ai-se/where.

{\em Validity} refers to the extent to which a piece of research actually
investigates what the researcher purports to investigate~\cite{SSA15}.
{\em Internal validity} checks if the differences found in
the treatments can be ascribed to the treatments under study. 

One threat to internal validity of our experiments is the choice
of {\em training and testing} data sets discussed in 
\fig{systems}. Recall that, while all our learners used the same
{\em testing} data set, our untuned learners were only given
access to {\em training} data.

Another threat to internal validity  is {\em instrumentation}. The very low $\mu$ and $\sigma$ error values
reported in this study are so small that it is reasonable to ask whether they are due to some instrumentation
quirk, rather than due to using a clever sample strategy:
\begin{itemize}
\item
Our low $\mu$ values are consistent with prior work~\cite{sarkar2015cost};
\item
As to our low $\sigma$ values, we note that, when the  error values are so close to 0\,\%, the standard
deviation of the error is ``squeezed'' between zero and those errors. Hence, we would expect that
experimental rigs
that generate error values on the order of 5\,\% and \eq{err} should have $\sigma$ values of $0\le \sigma \le 5$ (e.g., like those seen in our introduction).
\end{itemize}

Regarding SQLite, we cannot measure all possible configurations in reasonable time. Hence, we sampled only 100 configurations to compare prediction and actual performance values. We are aware that this evaluation leaves room for outliers.
Also, we are aware that measurement bias can cause false interpretations~\cite{me12d}. Since we aim at predicting performance for a special workload, we do not have to vary benchmarks.

  We aimed at increasing the {\em external validity} by choosing software systems from different domains with different configuration mechanisms and implemented with different programming languages. Furthermore, our subject systems  are deployed and used in the real world. Nevertheless, assuming the evaluations to be automatically transferable  to all configurable software systems is not fair. To further strengthen external validity, we run the model (generated by \textit{\what + $S_1$}) against other optimizers, such as NSGA-II and differential evolution~\cite{storn1997differential}. That is, we validated whether the learned models are not only applicable for GALE style of perturbation. In Table~\ref{fig:external_validity}, we see that the models developed are valid for all optimizers, as all optimizers are able to find the near optimal solutions.

\section{Related Work}
\label{sect:related}
 
In 2000, Shi and Maik~\cite{shi2000normalized} claimed the term ``spectral clustering'' as a reference to their normalized cuts
image
segmentation algorithm that  partitions data through a spectral (eigenvalue) analysis of the  
Laplacian representation of the similarity graph between instances in the data.

In 2003, Kamvar et al.~\cite{kamvar2003spectral}  generalized that definition saying that ``spectral learners''
were any data-mining algorithm that first replaced the raw
dimensions with those inferred from the spectrum (eigenvalues) of the affinity (a.k.a.\ distance)
matrix of the data, optionally adjusted via some normalization technique).

Our clustering based on first principal component splits the data on a   approximation to an eigenvector, found at each recursive level
of the data (as described in \tion{spect}). 
Hence, this  method is a ``spectral clusterer'' in the general Kamvar sense. 
Note that,
for our data, we have
not found that Kamvar's normalization matrices are needed.

Regarding sampling, there are a wide range of methods know as experimental designs or designs of experiments~\cite{pukelsheim2006optimal}. They usually rely on fractional factorial designs as in the combinatorial testing community~\cite{Kuhn:2013}. 

Furthermore, there is a recent approach that learns {\em per\-for\-mance-influence models} for configurable software systems~\cite{SGA+15}. While this approach can handle even numeric features, it has similar sampling techniques for the Boolean features as reported in their earlier work~\cite{siegmund2012predicting}. Since we already compared to that earlier work and do not consider numeric features, we did not compare our work to performance-influence models.

\section{Conclusions \& Future Work}

Configurable software systems today are widely used in practice, but they impose challenges
regarding finding performance-optimal configurations. State-of-the-art approaches require too
many measurements or are prone to large variances in their performance predictions. To overcome
these limitations, we have proposed a fast spectral learner, called \what,  along with three
new sampling techniques. The key idea of \what is to explore the configuration space with
eigenvalues of the features used in a configuration to determine exactly those configurations
for measurement that reveal key performance characteristics. 
This way, we can study many closely associated configurations with only a few measurements.

We evaluated our approach on six real-world configurable software systems borrowed from the
literature. Our approach achieves similar to lower error rates, while being stable when
compared to the state of the art. In particular, with the exception of Berkeley DB, our
approach is more accurate than the state-of-the-art approaches by Siegmund et
al.~\cite{siegmund2012predicting} and Guo et al.~\cite{guo2013variability}. Furthermore, we
achieve a similar prediction accuracy and stability as the approach by Sarkar et
al~\cite{sarkar2015cost}, while requiring a far smaller number of configurations to be
measured. We also demonstrated that our approach can be used to build cheap and stable
surrogate prediction models, which can be used by off-the-shelf optimizers to find the
performance-optimal configuration.  We use the correlation dimension to demonstrate how the high dimensional configuration space of our subject systems has a low intrinsic dimensionality, which might be the reason why \what performs so well on these datasets. 

As to future work, we plan to explore the implications of \what{}. Currently \what{} uses a static number of evaluations based on the total number of possible configurations ($\sqrt{N}$), which may not be useful for systems that are more difficult to model that the system used in this study. Hence, we need a progressive strategy, which can progressively sample new configurations and stop the sampling process based on either the performance score achieved or the budget allocated. Finally the current version of \what{}, assumes that all the features are of similar importance and uses a Euclidean distance to differentiate between good and `not-so-good' solutions. There are certainly systems where not all the features are equally important or there are redundancy in terms of configuration options. Hence, using feature weighting techniques to find weight (importance) of configuration options and use that information to differentiate between configurations.


\begin{acknowledgement}
The work is partially funded by NSF awards \#1506586. Sven Apel's work has been supported by the German Research Foundation (AP 206/4 and AP 206/6). Norbert Siegmund's work has been supported by the German Research Foundation (SI 2171/2).
\end{acknowledgement}

\bibliographystyle{plain}


\end{document}